%% file: main.tex
\documentclass[sigconf,9pt]{acmart}

\usepackage[english]{babel}
\usepackage{blindtext}

\renewcommand\footnotetextcopyrightpermission[1]{} 
\setcopyright{none}
\copyrightyear{2023}
\acmYear{2023}
\setcopyright{acmlicensed}\acmConference[This work is accepted at ACM SIGCOMM 2023]{ACM SIGCOMM 2023 Conference}{September 10--14, 2023}{New York, NY, USA}
\acmBooktitle{ACM SIGCOMM 2023 Conference (ACM SIGCOMM '23), September 10--14, 2023, New York, NY, USA}
\acmPrice{15.00}
\acmDOI{10.1145/3603269.3604857}
\acmISBN{979-8-4007-0236-5/23/09}



\usepackage{graphicx,subcaption}
\usepackage{amsmath}
\usepackage{xspace}
\usepackage{comment}
\usepackage{booktabs}
\usepackage{enumitem}
\usepackage{microtype}
\usepackage[title]{appendix}
\graphicspath{{figs/}}

\newcommand{\sysname}{\textsc{Teal}\xspace}

\usepackage[colorinlistoftodos,prependcaption,textsize=tiny]{todonotes}

\DeclareMathOperator*{\argmin}{arg\,min}

\newcommand{\parab}[1]{\smallskip\noindent\textbf{#1}~}
\newcommand{\parai}[1]{\smallskip\noindent\textit{#1}~}
\providecommand{\myparab}[1]{\smallskip\noindent\textbf{#1}~}

\providecommand{\eg}{e.g., }

\begin{document}

\newcommand{\authdag}{\textsuperscript{\textdagger}}
\newcommand{\authsec}{\textsuperscript{\textsection}}
\newcommand{\authp}{\textsuperscript{\textparagraph}}

\title[\sysname: Traffic Engineering Accelerated by Learning]{\sysname: Learning-Accelerated Optimization of \mbox{WAN Traffic Engineering}}

\subtitle{\vspace{-8pt}}

\author{Zhiying Xu}
\affiliation{%
  \institution{Harvard University}
  \country{}
}
\author{Francis Y. Yan}
\affiliation{%
  \institution{Microsoft Research}
  \country{}
}
\author{Rachee Singh}
\affiliation{%
  \institution{Cornell University}
  \country{}
}
\author{Justin T. Chiu}
\affiliation{%
  \institution{Cornell University}
  \country{\vspace{13pt}}
}
\author{Alexander M. Rush}
\affiliation{%
  \institution{Cornell University}
  \country{\vspace{13pt}}
}
\author{Minlan Yu}
\affiliation{%
  \institution{Harvard University}
  \country{\vspace{13pt}}
}

\renewcommand{\shortauthors}{Z. Xu, F. Yan, R. Singh, J. Chiu, A. Rush, M. Yu}

\date{}

\input{abstract}

\maketitle

\input{intro}

\input{motivation}

\input{design}

\input{implementation}

\input{eval}

\input{relwork}

\input{conclusion}

\section*{Acknowledgments}
\label{sec:acknowledge}
We thank our anonymous reviewers for their insightful comments
and Zhizhen Zhong for shepherding this work.
We also thank Srikanth Kandula, Firas Abuzaid, Yang Zhou, 
Deepak Narayanan, Fiodar Kazhamiaka, Umesh Krishnaswamy, and Victor Bahl
for their helpful feedback. 
This work was supported in part by the NSF grant CNS-2107078.

\newpage
\bibliographystyle{plain}
\bibliography{references}

\input{appendix}

\end{document}

%% file: abstract.tex
\begin{abstract}
The rapid expansion of global cloud wide-area networks (WANs) has posed a
challenge for commercial optimization engines to efficiently solve network
traffic engineering (TE) problems at scale. Existing acceleration strategies
decompose TE optimization into concurrent subproblems but
realize limited parallelism due to an inherent tradeoff between run time and
allocation performance.

We present \sysname, a learning-based TE algorithm that leverages the parallel
processing power of GPUs to accelerate TE control. First, \sysname designs a
flow-centric graph neural network (GNN) to capture WAN connectivity and
network flows, learning flow features as inputs to downstream allocation.
Second, to reduce the problem scale and make learning tractable, \sysname
employs a multi-agent reinforcement learning (RL) algorithm to independently
allocate each traffic demand while optimizing a central TE objective.
Finally, \sysname fine-tunes allocations with ADMM (Alternating Direction
Method of Multipliers), a highly parallelizable optimization algorithm for
reducing constraint violations such as overutilized links.

We evaluate \sysname using traffic matrices from Microsoft's WAN. On a large WAN
topology with $>$1,700 nodes, \sysname generates near-optimal flow allocations
while running several orders of magnitude faster than the production
optimization engine. Compared with other TE acceleration schemes, \sysname
satisfies 6--32\% more traffic demand and yields 197--625$\times$ speedups.
\end{abstract}

%% file: intro.tex
\section{Introduction}
\label{sec:intro2}
Large cloud providers invest billions of dollars to provision
and operate planet-scale wide-area networks (WANs) that interconnect
geo-distributed cloud datacenters. Cloud WANs play a vital role
in the operations of cloud providers as they enable low-latency and high-throughput
applications in the cloud. Over the last decade,
cloud providers have implemented
centralized network traffic
engineering (TE) systems based on SDN (software-defined networking)
to efficiently utilize
their cloud WANs~\cite{b4, swan, b4after, blastshield}.

\begin{figure}[t]
    \centering
    \includegraphics[width=\columnwidth]{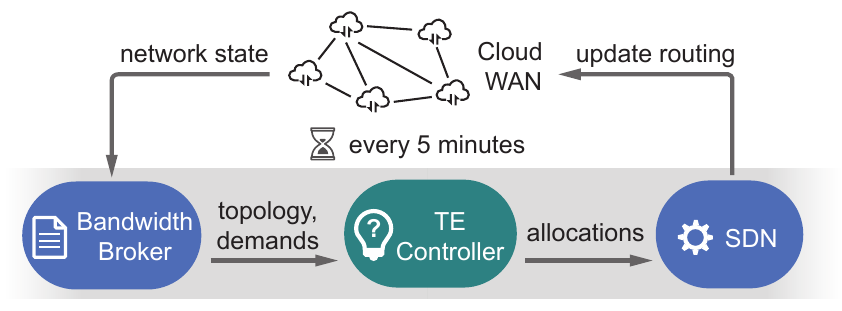}
    \vspace{-13pt}
    \caption{Control loop of WAN traffic engineering.}
    \label{fig:te_workflow}
\end{figure}

TE systems allocate demands between datacenters
to achieve high link utilization~\cite{swan, b4}, fairness among
network flows~\cite{swan}, and resilience to link failures in WANs~\cite{ffc, teavar}.
Traditionally, cloud WAN TE systems have approached
traffic allocation as an optimization problem,
with the objective of achieving a desired network property
(\eg minimum latency, maximum throughput).
To this end, they implement a software-defined TE controller
as illustrated in Figure~\ref{fig:te_workflow}. The TE controller
periodically (e.g., every five minutes) receives traffic demands
to allocate gauged by a bandwidth broker,
solves the TE optimization problem, and translates the traffic
allocations into router configurations to deploy through SDN.

After a decade of operation, production WAN TE systems are facing two major
challenges. First, the deployment of new edge sites and datacenters has
increased the size of cloud WANs by an order of magnitude~\cite{swan}.
Larger WAN topologies have increased the complexity of TE optimization
and the time required to solve it.
During the computation of
updated flow allocations (even when the five-minute time budget is
not exceeded), stale routes will remain in use and lead to suboptimal
network performance~\cite{ncflow}.
Second, WANs have evolved from carrying first-party discretionary traffic
to real-time and user-facing traffic~\cite{onewan}. As a result, cloud TE systems
must react to rapid changes in traffic volume, which is a hallmark of organic
user-driven demands.
Sudden topology changes due to link failures further exacerbate the negative effects
of long TE control on network performance.
Therefore, fast computation of traffic allocations is critical for TE systems
to retain performance on large WAN topologies.

While linear programming (LP) solvers used by TE systems
can find optimal solutions, they struggle to scale with the
growing network size.
State-of-the-art algorithms designed for accelerating TE optimization
address this challenge by decomposing the original problem
into smaller subproblems (through the partition of WAN
topology~\cite{ncflow} or traffic demands~\cite{pop}), and solve
them in parallel using LP solvers.
However, these algorithms face a fundamental tradeoff between speed
and performance in the decomposition, restricting themselves to
only \textit{a few dozen} subproblems and thus limited
parallelism (\S\ref{sec:scaling-te}).

Our key insight is that deep learning-based TE schemes may
unlock massive parallelism by utilizing \textit{thousands of} GPU
threads that are made readily accessible through modern deep learning
frameworks~\cite{paszke2019pytorch,tensorflow,chetlur2014cudnn}.
The enormous parallelism is owing to the well-known affinity between
neural networks and GPUs (e.g., SIMD operations on GPUs speed up matrix
multiplication), as well as the tremendous community efforts
for optimizing neural network
inference~\cite{jia2014caffe,inference2015performance,onnx}.
Meanwhile, by capitalizing on a wealth of historical data from production
WANs and exploiting traffic patterns,
learning-based algorithms are poised to simultaneously retain
TE performance as well.

Unfortunately, off-the-shelf deep learning models do not directly apply
to TE. First, standard fully-connected neural
networks fail to take into account the effects of WAN connectivity on traffic
allocations, yielding solutions that are far from optimal.
Second, the escalating scale of the TE problem makes it intractable to train a
monolithic model to navigate the high-dimensional solution space.
Finally, neural networks are unable to enforce constraints,
leading to unnecessary traffic drops due to exceeded link
capacities.

To address these challenges, we present a learning-accelerated TE scheme
named \sysname.
First, \sysname constructs a flow-centric graph neural network (GNN)
to capture WAN topology and extract informative features from traffic flows
for the downstream allocation task.
Next, \sysname allocates each demand individually using a shared
policy (neural) network
based on the learned features.
Doing so reduces the problem scale from global
traffic allocation to per-demand tasks, making the learning process more tractable
(in a low-dimensional space) and feasible (fit into GPU memory).
To coordinate the independent allocations of demands and avoid
contention for links, \sysname leverages multi-agent reinforcement
learning (RL) to train the end-to-end model---GNN and policy
network---toward
optimizing a central TE objective.
Finally, \sysname fine-tunes the model's output allocations using
a highly parallelizable constrained optimization
algorithm---ADMM (alternating direction method of multipliers),
which is well suited for reducing constraint violations
such as oversubscribed links and enhancing solution quality.

We evaluate \sysname on traffic matrices collected over a 20-day period
from Microsoft's WAN (\S\ref{sec:eval}).
Our experimental results show that on large WAN topologies,
\sysname realizes near-optimal flow allocation while
being several orders of magnitude faster than the production
TE optimization engine using LP solvers.
Compared with the state-of-the-art schemes for TE
acceleration~\cite{ncflow,pop,namyar2022minding}
on a large topology with $>$1,700 nodes,
\sysname satisfies 6--32\% more traffic demand and yields
197--625$\times$ speedups.
To aid further research and development, we have released \sysname's source code
at \url{https://github.com/harvard-cns/teal}.

%% file: motivation.tex
\section{Background and Motivation}
\label{sec:motivation}

Production WANs rely on a centralized TE controller to
allocate traffic demands between datacenters, which are gauged by
a bandwidth broker periodically (e.g., every 5 minutes).
The TE controller splits the traffic demand onto a handful of precomputed paths
(e.g., 4 shortest paths~\cite{ncflow,pop}) between the demand's source and
destination, with the goal of maximizing a TE objective (e.g., overall
throughput) while satisfying a set of constraints (e.g., link
capacities). This path formulation of TE is widely adopted in
production inter-datacenter WANs~\cite{swan,b4,b4after,blastshield}.
At its core, TE optimization is a multi-commodity flow problem
(formally defined in Appendix~\ref{sec:te-formulation}),
which is traditionally solved with linear programming (LP) solvers.
In this section, we begin with the scalability crisis faced by
today's TE systems, and motivate the need
and challenges for a learning-accelerated TE solution.

\subsection{Scaling challenges of TE}
\label{sec:scaling-te}

\begin{figure}[t]
    \centering
    \includegraphics[width=0.75\columnwidth]{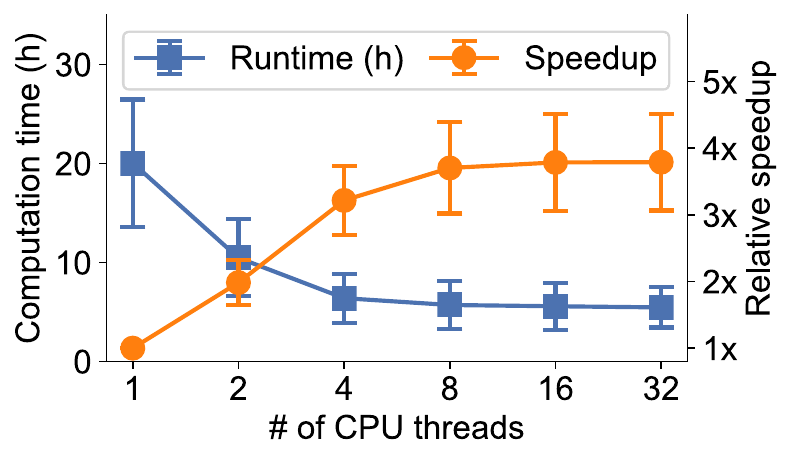}
    \vspace{-4pt}
    \caption{On a topology with $>$1,700 nodes (ASN in
    Table~\ref{tab:topologies}), the TE optimization using the Gurobi solver
    experiences a marginal speedup as more CPU threads become available.}
    \label{fig:cpu-asn}
\end{figure}

In their early years, cloud WANs only consisted of tens of datacenters, so it
was feasible for commercial LP solvers to compute traffic allocations
frequently. However, the rapid deployment of new datacenters has
rendered the TE task prohibitively slow at scale, requiring hours for commercial
solvers to allocate traffic on WANs with thousands of nodes. Consequently,
WAN operators are seeking to accelerate TE optimization to keep pace with the
growing size of the WAN.

\parab{Parallelizing LP solvers.}
An intuitive way of accelerating TE optimization is to parallelize
state-of-the-art LP solvers, such as Gurobi~\cite{gurobi} and
CPLEX~\cite{cplex}.
Figure~\ref{fig:cpu-asn} evaluates the speedup of the Gurobi solver
on a WAN topology with more than 1,700 nodes
(ASN in Table~\ref{tab:topologies}). As more CPU threads are made available,
we observe that the speedup is sublinear and marginal. E.g., using 16 threads
only makes Gurobi $3.8\times$ faster, which still requires 5.5 hours to
complete a flow allocation.
This is due to LP solvers' sequential nature, e.g.,
the conventional simplex
method~\cite{simplex} takes one small step at a time toward the optimal
solution along the edges of the feasible region, and requires thousands to
millions of such steps to complete.
To exploit multiple CPU threads,
LP solvers often resort to concurrently running \textit{independent}
instances of different optimization algorithms~\cite{parallel-gurobi},
where each instance executes serially on a separate thread;
the solution is yielded by whichever instance completes first.
This is apparently not an efficient use of CPU capacity, thus resulting in
marginal speedups on multiple CPUs.

\parab{Approximation algorithms.} Combinatorial algorithms, such as
the Fleischer's algorithm~\cite{fleischer2000approximating}, are designed to compute approximate
but asymptotically faster solutions to the underlying network flow problem
of TE. Despite having a lower time complexity than LP solvers in theory,
these approximation algorithms are found to be hardly faster in
practice~\cite{ncflow}. The reason is that these algorithms remain
\textit{iterative} in nature, incrementally allocating more flows until the 
solution quality is deemed adequate (yet suboptimal), which often results in an
excess of iterations to terminate.

\parab{Decomposing TE into subproblems.} Recently,
NCFlow~\cite{ncflow} and POP~\cite{pop} introduced techniques to
decompose TE optimization into subproblems, applying LP solvers
simultaneously in each subproblem and merging their results at the end.
NCFlow partitions the network spatially into $k$ clusters,
whereas POP creates $k$ replicas of the network, each with $1/k$
link capacities, and randomly assigns traffic demands to these replicas.
Although a larger $k$ reduces the overall run time,
it also fundamentally impairs the TE performance.
Moreover, NCFlow also requires nontrivial coordination during the merging process.
Consequently, NCFlow and POP are forced to adopt small values of $k$
(e.g., 64--81 on a network of 754 nodes).
In \S\ref{sec:eval}, we show that NCFlow and POP are substantially
slower than our learning-accelerated approach, while having
notably worse allocation performance.

\subsection{Accelerate TE optimization with ML}
\label{sec:accelerating-te}

To cope with the growing scale of TE,
we argue that with judicious design, machine learning (ML) can significantly
accelerate TE optimization.
By training on a vast amount of historical traffic data,
ML-based TE schemes also have the potential to attain
near-optimal allocation performance.

\myparab{Unlocking massive parallelism.}
Encoding a TE algorithm in neural networks transforms the traditionally
iterative TE optimization (LP solvers or combinatorial algorithms)
into the inference process of neural networks, where the input data
(e.g., traffic demands on a WAN topology) is propagated
in the forward direction through the neural network to compute the output
(e.g., traffic splits on the preconfigured paths).
This inference process unlocks massive parallelism due to mainly consisting
of highly parallelizable operations such as matrix multiplications.

\myparab{Leveraging hardware acceleration.}
Modern deep learning frameworks~\cite{paszke2019pytorch,tensorflow,onnx} 
have empowered neural networks to easily leverage \textit{thousands of} threads
on GPUs (or other specialized hardware~\cite{tpu}). They can greatly accelerate
the computation of learning-based TE systems compared
with state-of-the-art schemes~\cite{ncflow,pop},
which are fundamentally limited to \textit{tens of} parallel workers.
In addition, the deep learning community has integrated various
optimization
techniques~\cite{jia2014caffe,inference2015performance,onnx}
into these frameworks, further accelerating neural network inference.

\myparab{Exploiting abundant training data.}
Operational WANs generate an abundance of traffic data
that can be used to train neural networks. A carefully designed ML-based TE
scheme is capable of discovering regularities in the training
data---such as patterns in graph connectivity, link capacities, and 
traffic demands---and ultimately learns to optimize allocation performance with respect
to operator-specified objectives.

\subsection{Challenges of applying ML to TE}
\label{sec:ml-te-challenges}

While holding the promise of near-optimal performance
and substantial speedup relative to LP-based TE methods,
deep learning is not a panacea. In fact,
using ML for TE optimization is not as straightforward as it may appear.

\parab{Graph connectivity and network flows.} 
Naively using vanilla fully-connected neural networks for TE optimization
would ignore the connectivity in WAN topology.
While graph neural networks (GNNs)~\cite{gentle-gnns,wu2020comprehensive},
designed for graph-structured input data, can model traditional graph attributes
such as nodes and edges, their unmodified form is inadequate to model
network \textit{flows}---the focal point of TE optimization.

\parab{High-dimensional solution space.}
In the path formulation of TE widely adopted in practice
(details in Appendix~\ref{sec:te-formulation}), the TE controller splits
each demand across a handful of preconfigured
paths, such as 4 shortest paths. Therefore, representing the flow 
allocation for a topology of $N$ nodes requires
$O(N^2)$ split ratios.
To put it into context, on a topology with 1,000 nodes, the solution
space would contain up to 4 million dimensions, exposing ML-based TE methods
to the ``curse of dimensionality''~\cite{koppen2000curse}.

\parab{Constrained optimization.}
Unlike LP solvers, neural networks are known to lack the capability to
enforce constraints on outputs~\cite{lee2017enforcing}.
As a result, the traffic allocations generated by neural networks
may exceed certain link capacities when deployed directly,
leading to network congestion and reduced TE performance.

To tackle the above challenges, we propose the following designs:
\textit{i)} a flow-centric GNN (called ``FlowGNN'') to capture
WAN connectivity and model network flows (\S\ref{sec:flow-gnn});
\textit{ii)} a multi-agent reinforcement learning (RL) algorithm that allocates
each traffic demand independently to reduce the problem scale and make
learning tractable (\S\ref{sec:marl});
\textit{iii)} solution fine-tuning using the alternating direction method
of multipliers (ADMM) to minimize constraint violations such as
link overutilization (\S\ref{sec:admm}).

%% file: design.tex
\section{\sysname: Learning-accelerated TE}
\label{sec:design}

In this section, we present the design of \sysname---\textbf{T}raffic
\textbf{E}ngineering \textbf{A}ccelerated with \textbf{L}earning.
The goal of \sysname is to train a fast and scalable TE scheme
through deep learning while achieving near-optimal traffic allocation
on large-scale topologies. The rationale behind using deep learning is to harness the massive parallelism and hardware acceleration unlocked by
neural networks. Moreover, every component of \sysname is carefully designed
to be \textit{parallelizable} (fast on GPUs) and \textit{scalable}
(performant on large WANs).

\subsection{Overview}
\label{sec:teal-overview}

\begin{figure}[t]
    \centering
    \includegraphics[width=0.9\columnwidth]{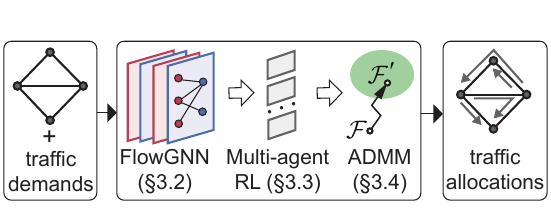}
    \caption{Workflow of \sysname. \sysname inputs traffic demands and
    link capacities into FlowGNN to learn flow embeddings (\S\ref{sec:flow-gnn}),
    which are then mapped to initial traffic allocations through
    multi-agent RL (\S\ref{sec:marl}).
    ADMM subsequently fine-tunes the allocations and mitigates
    constraint violations (\S\ref{sec:admm}).}
    \label{fig:teal-design}
\end{figure}

We begin by outlining the workflow of \sysname during deployment
(Figure~\ref{fig:teal-design}). Upon the arrival of a new traffic demand matrix or
a change in link capacity\footnote{We note that link failures
can be viewed as an extreme scenario of capacity change, where the capacity of
a failed link is reduced to zero.}, \sysname passes the updated traffic
demands and current link capacities into a novel graph neural network (GNN)
that we call \textit{FlowGNN}~(\S\ref{sec:flow-gnn}). FlowGNN learns to 
transform the demands into compact feature vectors known as ``embeddings,'' 
which preserve the graph structure and encode the flow information required for
the downstream allocation. These flow embeddings are extracted by FlowGNN in a
parallel and parameter-efficient manner that scales well with the size of
the WAN topology.

In the widely adopted path formulation of TE
(details in Appendix~\ref{sec:te-formulation}),
each traffic demand is split into multiple flows over a set of
preconfigured paths (e.g., 4 shortest paths~\cite{ncflow,pop}).
To determine the split ratios of a given demand,
\sysname aggregates the embeddings learned for each flow 
of the demand and inputs them into a shared policy (neural) network. 
The policy network, which allocates demands independently, 
is trained offline to coordinate flow allocations toward optimizing a global
TE objective (e.g., total flow),
through a \textit{multi-agent reinforcement learning (RL)} algorithm
we customize for TE (\S\ref{sec:marl}).
This design enables processing demands individually rather than the entire
traffic matrix at once, making the policy network more compact (in terms of the 
parameters to learn) and oblivious to the WAN topology size.

So far, the split ratios output by the policy network might still
exceed certain link capacity constraints, resulting in dropped traffic
and suboptimal TE performance.
To mitigate constraint violations and enhance the solution quality,
\sysname augments the neural networks (FlowGNN and policy network) with 2--5 rapid iterations of a
classical constrained optimization
algorithm---\textit{alternating direction method of multipliers (ADMM)}
(\S\ref{sec:admm}). During each iteration, ADMM starts from a potentially
infeasible TE solution with capacity violations and advances
toward the feasible region,
fine-tuning traffic splits to meet more constraints and
improve the overall TE performance.
Each iteration of ADMM is inherently parallel.
When warm-started with a reasonably good solution such as the one
generated by the neural networks, ADMM can attain a noticeable
improvement in performance within several fast iterations.

For each WAN topology, \sysname trains its ``model''---FlowGNN and policy
network---end to end to optimize an operator-provided TE objective; ADMM
requires no training. All the three key components of \sysname
(FlowGNN, multi-agent RL, and ADMM) are carefully designed
to be highly parallelizable,
enabling fast computation and scalable TE performance as the size of the WAN
topology grows.

\subsection{Feature learning with FlowGNN}
\label{sec:flow-gnn}

In light of the graph-based structure of WAN topologies, \sysname leverages
graph neural networks (GNNs) for feature learning.
GNNs are a family of neural networks designed to
handle graph-structured data~\cite{gentle-gnns}
and have found applications in various domains, including
network planning~\cite{zhu2021network},
social network~\cite{fan2019graph, mohamed2020social},
and traffic prediction~\cite{lange2020traffic}.

GNNs typically store information in graph attributes, commonly in nodes,
using a compact vector representation known as
\textit{embeddings}~\cite{hamilton2017representation}.
To preserve graph connectivity in the embeddings,
neighboring nodes in the GNN exchange information through
``message passing''~\cite{message-passing}:
Each node collects the embeddings from adjacent nodes,
transforms the aggregated embeddings using a learned transformation function
(e.g., encoded in a fully-connected neural network),
and updates its own embedding with the result.
GNNs are intrinsically parallel as message passing occurs 
simultaneously across nodes.
Applying message passing once constitutes one GNN layer,
and stacking multiple GNN layers allows
information to be disseminated multiple hops away.
It is noteworthy that GNNs are parameter efficient because each layer shares
the same transformation function that operates in
the low-dimensional embedding space, which does not grow in proportion
to the input graph size.

Despite the strengths of GNNs, the primary focus of TE is the assignment of 
\textit{flows}, in which each flow originating from a demand follows a
predefined \textit{path} along a chain of network links (\textit{edges}).
TE is concerned with the interference between flows as they compete
for link capacities.
Hence, we put a spotlight on flows and explicitly represent
flow-related entities---edges and paths---as the nodes in our
TE-specific GNN, which we call \textit{FlowGNN}.

\begin{figure}[t]
    \centering
    \includegraphics[width=0.8\columnwidth]{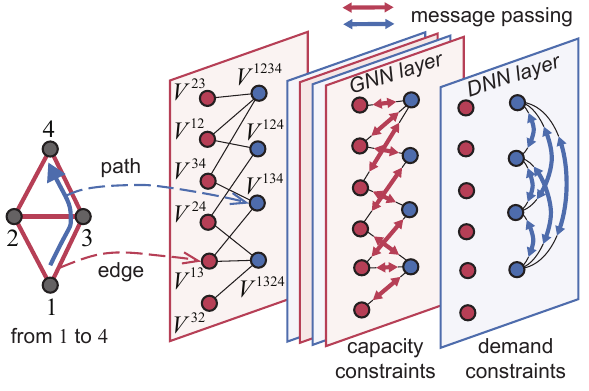}
    \caption{Illustration of a FlowGNN construction.
    FlowGNN alternates between GNN layers that are designed to capture
    capacity constraints,
    and DNN layers that are intended to capture demand constraints.}
    \label{fig:flow-gnn}
\end{figure}

Figure~\ref{fig:flow-gnn} exemplifies the construction of a FlowGNN.
At a high level, FlowGNN alternates between
\textit{GNN layers} aimed at capturing capacity constraints,
and \textit{DNN layers} aimed at capturing demand constraints,
which dictate that the total volume of all flows derived from
a demand should not exceed the demand itself
(formal formulation in Appendix~\ref{sec:te-formulation}).

The GNN layer in FlowGNN is structured as a bipartite graph.
For each edge in the input topology, we create an ``EdgeNode''
(e.g., $V^{13}$ for the edge connecting nodes \#1 and \#3),
and for each preconfigured path associated with a demand,
we create a ``PathNode''
(e.g., $V^{134}$ for the path containing nodes \#1, \#3, and \#4).
An EdgeNode and a PathNode are connected in the GNN layer if
and only if the edge lies on the path
(e.g., $V^{13}$ is connected to $V^{134}$ but not to $V^{124}$).
The intuition of this setup is to allow EdgeNodes and PathNodes
to interact and learn to respect capacity constraints during
message passing. For example, when an edge serves as a bottleneck
for competing flows, the EdgeNode's embedding will be
influenced by its neighboring PathNodes.
During initialization, the embedding of an EdgeNode
is initialized with the capacity of the corresponding edge,
(e.g., $V^{13}$ is initialized to the link capacity
between nodes \#1 and \#3),
while the embedding of a PathNode
is initialized with the volume of the associated demand
(e.g., $V^{134}$ is initialized to the traffic demand
from node \#1 to node \#3).
In doing so, a PathNode's embedding becomes dependent on the
corresponding demand specified in a traffic matrix, thereby
capturing a \textit{flow} routed on the path
(rather than the underlying physical network path).

Due to the absence of connections between PathNodes, 
the GNN layer is unable to make each PathNode aware of the other PathNodes
associated with the same demand.
To address this, we add a DNN layer after each GNN layer
to coordinate flows---represented by their PathNodes---that
stem from the same demand.
The DNN layer, a fully-connected neural network, essentially
transforms and updates the embeddings of the related PathNodes
(e.g., $V^{1234}$, $V^{124}$, $V^{134}$, and $V^{1324}$ for the demand
from node \#1 to node \#4).
Specifically, these embeddings
are fed into the DNN layer
to obtain an equal number of updated embeddings, which are then stored back into
the respective PathNodes.

Once the FlowGNN is fully trained (in conjunction with the policy network
described next), it learns to generate embeddings that encode
the graph-structured input of TE in the embedding space.
In particular, the final embeddings associated with PathNodes
represent the learned feature vectors of flows traversing those paths
and serve as informative input for the following task of
flow allocation. We visualize the learned flow
embeddings in \S\ref{sec:visualization} to interpret
their encoded knowledge about path congestion.

\subsection{Flow allocation with multi-agent RL}
\label{sec:marl}

Given the flow embeddings generated by FlowGNN as feature inputs,
\sysname creates a \textit{policy network}
to map these embeddings to traffic splits on the corresponding paths,
materializing flow allocation. 
The FlowGNN and policy network constitute the ``model'' of \sysname,
which is trained end to end to optimize an operator-specified TE objective.

Since a network link is frequently utilized by many competing flows,
an ideal policy network should process all flows simultaneously to
determine the globally optimal allocations.
However, this approach entails enormous input and output spaces,
resulting in a gigantic neural network with a large number of parameters.
To put it in perspective, for a WAN topology with
a thousand nodes, the ideal policy network would require
\textit{millions of} flow embeddings as input, and output an equal
number of split ratios, one for each flow.
In practice, we find that this type of gigantic policy network is
difficult to train and leads to a significant amount of demand unfulfilled
(\S\ref{sec:eval_interpret}).

\begin{figure}[t]
    \centering
    \includegraphics[width=0.7\columnwidth]{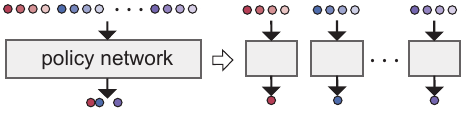}
    \caption{\sysname processes each demand independently using a shared,
    significantly smaller policy network.}
    \label{fig:distribute_policy}
\end{figure}

To reduce the problem dimension and the number of parameters to learn,
\sysname processes each demand independently\footnote{This
approach bears resemblance to distributed TE~\cite{kandula2005walking},
but we target a centralized TE controller with full visibility
into the entire WAN topology.}
using a shared policy network that is significantly smaller in size
(as illustrated in Figure~\ref{fig:distribute_policy}).
For instance, when assigning a traffic demand across four candidate paths,
our policy network only obtains four (low-dimensional) flow embeddings from FlowGNN
as input (e.g., the embeddings of $V^{1234}$, $V^{124}$, $V^{134}$,
and $V^{1324}$), and outputs four split ratios to prescribe the allocation.
Different demands are processed simultaneously as a batch
input to the policy network.
This design allows the policy network to be agnostic to the
WAN topology size, making it more compact and feasible to learn.

Despite the benefits, allocating each demand independently can result in
a lack of coordination unless the policy network is
trained---along with FlowGNN---to be aware of a central TE objective.
This raises the question:
\textit{What learning algorithm is suitable for training \sysname's model,
which generates local traffic splits for each demand while optimizing
a global TE objective?}
To address this question, we discuss several candidates
below before landing on multi-agent RL.

\parab{Supervised learning:} In an offline setting, LP solvers such as
Gurobi can be used to compute the optimal traffic allocations,
thereby providing ground-truth traffic splits for
\sysname's model to learn using standard supervised learning.
Nevertheless, generating these ground-truth labels for large WANs
can be excessively time-consuming and incur substantial memory usage.
E.g., Gurobi requires 5.5~hours to find the optimal allocations for a single
traffic matrix on a 1739-node topology, while consuming up to 18.1~GB of memory.

\parab{Direct loss minimization:} Supervised learning minimizes
the distance between the optimal splits of each demand and the output
splits from \sysname's model as the loss.
In fact, any \textit{differentiable} loss function
can be used instead and minimized directly through gradient
descent, referred to as
``direct loss minimization''~\cite{hazan2010direct,song2016training}.
However, common TE objectives are \textit{non-differentiable}.
E.g., calculating the total (feasible) flow requires reconciling flows
that collectively exceed a link's capacity, such as by proportionally
dropping traffic from each flow.
Consequently, the gradient of the total feasible flow
with respect to model parameters is zero\footnote{In this case,
we also refer to the loss function as ``non-differentiable.''},
thus preventing learning through gradient descent.
To address this, a common workaround is to approximate the non-differentiable
loss function with a differentiable ``surrogate loss.''
In \S\ref{sec:eval_interpret}, we define a surrogate loss
that approximates the total feasible flow and implement a baseline
to directly minimize this surrogate loss.
However, using a surrogate loss entails approximation errors, and
identifying a suitable surrogate loss for a different TE objective may not
be straightforward.

\parab{Multi-agent RL:} \sysname opts for employing the framework of
multi-agent RL~\cite{foerster2016learning, vinyals2019grandmaster, coma},
casting the allocation of each demand
as an RL agent striving to attain a shared objective
in collaboration with other agents.
Each agent utilizes locally accessible information, i.e., the flow embeddings of
its own demand, to independently generate traffic splits.
During training, however, TE allows us to simulate the combined effect of
local traffic splits and compute a global TE objective, which serves as
an accurate signal, or ``reward,'' to guide the RL agents.
The desired TE objective (e.g., the non-differentiable total feasible flow)
can be used directly as the reward because RL algorithms do not require a 
differentiable reward function.
Upon the completion of training, each agent executes independently
without attending to the other agents.
This learning paradigm---centralized training of decentralized policies---is 
standard in the multi-agent setting when applicable,
with COMA~\cite{coma} being the state of the art.
We tailor COMA to TE and implement it in \sysname as follows.

Our variant of COMA, referred to as \textit{COMA$^*$},
is also based on policy gradients~\cite{sutton1999policy},
a workhorse of modern deep RL that
optimizes a parameterized policy (as encoded in our policy network)
with respect to the long-term return (expected cumulative reward).
Unlike COMA, our COMA$^*$ capitalizes on the fact that
the traffic allocations in TE computed for one time interval 
do not affect future intervals (e.g., traffic matrices).
This domain-specific insight allows us to improve training
by reducing the long-term return to a one-step return,
as well as enhance another key mechanism within COMA (related to
the estimation of the reward contributions of individual agents).
We include the details of COMA$^*$ in Appendix~\ref{sec:coma-details}.

\subsection{Solution fine-tuning with ADMM}
\label{sec:admm}
In essence, TE is a constrained optimization problem, yet neural networks
are known to be inadequate in enforcing constraints,
such as link capacities in TE.
As a result, the traffic allocations directly generated by \sysname's
neural network model are prone to link overutilization.
To mitigate constraint violations and enhance solution quality,
\sysname fine-tunes the allocation results using
2--5 fast iterations of
ADMM (Alternating Direction Method of Multipliers)~\cite{boyd-admm},
a classical constrained optimization algorithm. 

We outline the mechanism of ADMM below with additional details
provided in Appendix~\ref{sec:admm-details}.
ADMM is a variant of the augmented Lagrangian
method~\cite{bertsekas2014constrained},
which transforms the original constrained optimization problem into a series of
unconstrained problems by converting constraints into penalty terms 
in the objective function.
Applying ADMM in TE optimization requires that we
decouple constraints properly and introduce auxiliary variables
(a standard optimization technique) for each edge, path, or demand
based on the corresponding constraints.
Then, in each iteration, ADMM alternates between \textit{i)} minimizing the augmented
Lagrangian with respect to one variable while keeping other variables fixed,
and \textit{ii)} updating the variables in a manner that balances
optimization and constraints.

ADMM is well-suited to \sysname for two reasons.
First, unlike the widely used optimization methods in LP solvers,
such as simplex and interior-point methods, ADMM
does not require a constraint-satisfying solution to begin with.
Instead, ADMM allows starting from a constraint-violating point
(as \sysname's neural networks might output) and
iteratively moves toward the feasible region.
Second, ADMM is highly parallelizable because the minimization of the augmented
Lagrangian can be decomposed into many subproblems, each solving
for a single variable, e.g., created for each path or edge in TE.
These subproblems can be solved in parallel and benefit from
significant acceleration on GPUs.

Additionally, we note that using ADMM alone does not accelerate TE optimization.
This is because
when initialized randomly, ADMM still requires an excessive number of 
iterations to converge to an acceptable solution,
forfeiting its fast speed \textit{within} each iteration.
In contrast, \sysname's neural networks can warm-start ADMM
with a reasonably good solution, allowing ADMM to perform fine-tuning
and attain a noticeable improvement in several iterations with a negligible
impact on the overall run time.

%% file: implementation.tex
\section{Implementation of \sysname}
\label{sec:implementation}

\parab{Implementing \sysname.} 
In each time window (e.g., 5 minutes),
\sysname takes as input a WAN topology with link capacities,
a traffic matrix indicating the demand between every node pair,
and 4 precomputed paths to route each demand.
The output is 4 split ratios for each demand,
prescribing its allocation across the precomputed paths.
We implemented all three key modules of \sysname in PyTorch. Hyperparameters
(e.g., the number of neural network layers)
are tuned empirically by testing various values
(\S\ref{sec:eval_interpret}).

\begin{itemize}[noitemsep,topsep=0pt,leftmargin=*]

\item
\parai{FlowGNN.} FlowGNN comprises 6 GNN layers
interleaved with 6 DNN layers.
The embeddings in the first GNN layer are initialized as described in
\S\ref{sec:flow-gnn}, each with a single element.
In each of the following GNN layers, the embedding dimension is expanded by
one element, filled with the same value as the original initialization
(a technique to enhance the expressiveness of GNNs~\cite{nair2020solving}).
The final output embeddings consist of 6 elements each.

\item
\parai{Multi-agent RL.} The policy network in \sysname is
implemented as a fully-connected neural network with a single hidden
layer of 24 neurons.
It has 24 input neurons to receive 4 flow embeddings from FlowGNN
for each demand, and uses 4 output neurons, followed by a softmax
normalization, to generate 4 split ratios.

\item
\parai{ADMM.} We apply two iterations of ADMM for topologies with fewer than 100
nodes, and five iterations otherwise.

\end{itemize}

\parab{Training \sysname.} We train a separate \sysname model per WAN topology
and per TE objective. 
The Adam optimizer~\cite{kingma2014adam} is employed for
stochastic gradient descent, with a learning rate of $10^{-4}$.
Training \sysname from scratch takes approximately a week to complete on
large WAN topologies, such as ASN.

\parab{Retraining.} We retrain \sysname if a new node or link
is added to the WAN topology permanently.
We demonstrate in \S\ref{sec:eval_fail} that transient link failures do not
require retraining.
Compared with training from
scratch, each retraining session of \sysname only takes 6--10 hours.

%% file: eval.tex
\section{Evaluation}
\label{sec:eval}

In this section, we first describe our evaluation methodology in 
\S\ref{sec:methodology}.
Next, we compare \sysname with state-of-the-art TE schemes in
\S\ref{sec:comparison-baselines} and show that \sysname simultaneously achieves
substantial acceleration and near-optimal flow allocation.
\S\ref{sec:eval_fail} demonstrates \sysname's fast reaction to link failures,
and \S\ref{sec:robustness} assesses \sysname's robustness to temporal
and spatial fluctuations in traffic demands.
\S\ref{sec:eval-objectives} evaluates \sysname's flexibility
with respect to different TE objectives.
\S\ref{sec:offline-performance} reports the idealized offline performance
of TE schemes disregarding their computation times.
Finally, we examine the contributions of \sysname's individual components
in \S\ref{sec:eval_interpret}, and visualize its learned flow embeddings in \S\ref{sec:visualization} to interpret its behaviors.

\begin{table}[t]
  \centering
  \begin{tabular}{lll}
  \toprule
            & \# of nodes & \# of edges \\
  \midrule
  B4        & 12          & 38         \\
  SWAN      & $O(100)$    & $O(100)$    \\
  UsCarrier & 158         & 378         \\
  Kdl       & 754         & 1,790        \\
  ASN       & 1,739       & 8,558       \\
  \bottomrule
  \end{tabular}
  \vspace{10pt}
  \caption{Network topologies in our evaluation.}
  \label{tab:topologies}
  \vspace{-5pt}
\end{table}

\subsection{Methodology}
\label{sec:methodology}

\parab{Topologies.} We consider five WAN topologies:
Google's private WAN (B4~\cite{b4}),
Microsoft's software-defined WAN (SWAN~\cite{swan}),
two topologies from the Internet Topology
Zoo~\cite{knight2011internet}---UsCarrier and Kdl---and
an AS-level internet topology~\cite{caida_as} adapted for WAN purposes,
denoted as ``ASN.''
Their numbers of nodes and edges are summarized in Table~\ref{tab:topologies},
with additional topology characteristics in Appendix~\ref{sec:topology-details}.
We adopt the path formulation of TE used in production
(Appendix~\ref{sec:te-formulation}),
and precompute 4 shortest paths~\cite{ncflow,pop} between every pair of nodes
as the candidate paths to allocate flows.
In cases where link capacities are not provided,
we set the capacities to ensure that the best-performing TE scheme
satisfies a majority of traffic demand.

\parab{Traffic data.} We collect traffic data over a 20-day period in 2022
from SWAN, the production inter-datacenter WAN at Microsoft.
The total traffic observed in each 5-minute interval between
every source-destination pair is considered as their demand.
To translate these traffic demands from SWAN to other topologies,
we map each new node pair to a random node pair in SWAN,
and randomly sample disjoint sequences of traffic matrices,
with 700 consecutive intervals for training,
100 for validation, and 200 for testing.

\parab{Baselines.} We compare \sysname against the following baselines.

\begin{itemize}[noitemsep,topsep=0pt,leftmargin=*]

\item
\parai{LP-all:} LP-all solves the TE optimization problem for
\textit{all} demands using linear programming (LP).
Gurobi~\cite{gurobi} v9.1.2 is employed as the LP solver.

\item
\parai{LP-top:} LP-top implements a simple yet effective heuristic TE algorithm
that is recently revealed as ``demand pinning''~\cite{namyar2022minding}.
It allocates the top $\alpha \%$ of demands using an LP solver
and assigns the remaining demands to the shortest paths.
To balance allocation quality and computation time,
we set $\alpha=10$ after testing multiple values.
In our traffic trace, the top 10\% of demands account for a vast
majority (88.4\%)  of the total volume.

\item
\parai{NCFlow:} NCFlow~\cite{ncflow} partitions the topology into
disjoint clusters and concurrently solves the subproblem of TE optimization
within each cluster using an LP solver.
The results obtained from each cluster are then merged in a nontrivial
fashion to generate a valid global allocation.
We adopt the same number of clusters as specified in the paper for
UsCarrier and Kdl,
and apply the default partitioning algorithm (``FMPartitioning'') for
other topologies.

\item
\parai{POP:} POP~\cite{pop} replicates the entire topology $k$ times,
with each replica having $1/k$ of the original link capacities.
The traffic demands are randomly distributed to these replicas,
and each subproblem is solved in parallel with an LP solver.
We set $k$ based on the topology size,
with $k=1$ for B4 and SWAN, $k=4$ for UsCarrier, and $k=128$ for Kdl and ASN.
Client splitting threshold is set to $0.25$ to break down large demands.

\item
\parai{TEAVAR$^*$:} TEAVAR~\cite{teavar} is a TE scheme
that takes into account the risk of link failures.
It balances link utilization with
operator-defined availability requirements when allocating traffic.
We compare \sysname with TEAVAR$^*$, a variant of TEAVAR
adapted by NCFlow to maximize the total flow.

\end{itemize}

\begin{figure*}[t]
\centering
\begin{subfigure}[t]{0.48\textwidth}
\includegraphics[width=\textwidth]{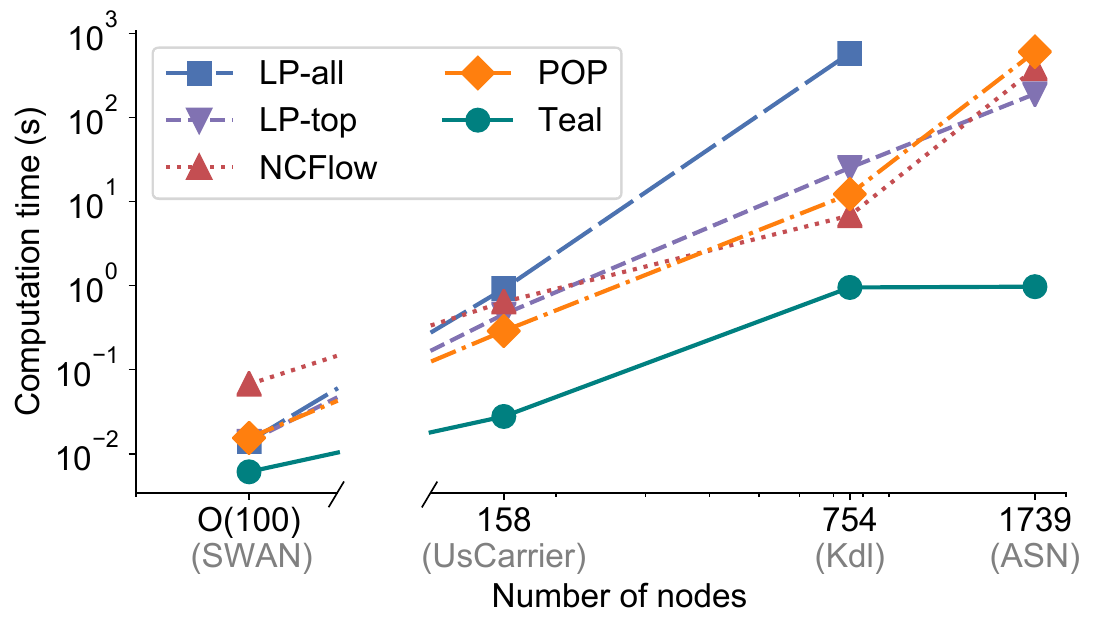}
\vspace{-10pt}
\caption{Average computation time}
\label{fig:basic_t}
\end{subfigure}
\begin{subfigure}[t]{0.48\textwidth}
\includegraphics[width=\textwidth]{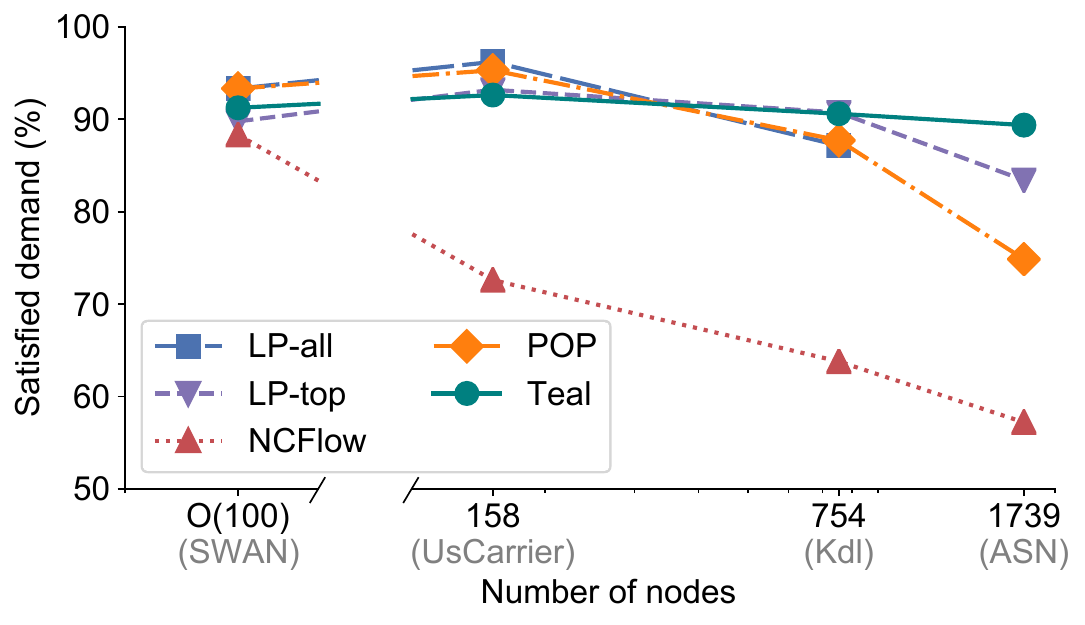}
\vspace{-10pt}
\caption{Average satisfied demand}
\label{fig:basic_d}
\end{subfigure}
\caption{Comparing \sysname with LP-all, LP-top, NCFlow, and POP
across different networks (LP-all is not feasible on ASN).
Designed to accelerate TE optimization on large topologies
such as Kdl and ASN, \sysname attains scalable performance as the
network size grows, reducing computation time to less than 2 seconds
while satisfying comparable or higher demand.}
\vspace{10pt}
\label{fig:basic}
\end{figure*}

\parab{Objectives.} Our default TE objective is to maximize the
total (feasible) flow~\cite{ncflow,swan,b4}.
Section~\ref{sec:eval-objectives} evaluates two additional
objectives: minimizing the max link utilization
(MLU)~\cite{fleischer2000approximating,learning-to-route},
and maximizing the total flow with delay penalties~\cite{mate}.

\parab{Metrics.} We consider the following
performance metrics.

\begin{itemize}[noitemsep,topsep=0pt,leftmargin=*]

\begin{table}[t]\centering
\begin{minipage}[b]{0.48\textwidth}
\centering
\begin{tabular}{ll}
\toprule
Algorithm & Computation time \\
\midrule
\sysname & Total run time (with GPU)\\
LP-all & Gurobi run time\\
LP-top &  Gurobi run time $+$ model rebuilding time\\
NCFlow  & Gurobi run time $+$ time to coalesce subproblems\\
POP & Gurobi run time\\
TEAVAR$^*$ & Gurobi run time\\
\bottomrule
\end{tabular}
\vspace{10pt}
\caption{Breakdown of computation time for each scheme.}
\label{tab:computation_time}
\end{minipage}
\vspace{-10pt}
\end{table}

\item
\parai{Computation time:}
We measure the total time required by each scheme to compute
flow allocation amortized over each traffic matrix, while carefully
excluding one-time costs such as the initialization time.
The measurement is conducted on 16 
CPU threads (Intel Xeon E5-2680) following the setup in
NCFlow~\cite{ncflow}. An additional GPU (Nvidia Titan RTX) is
made available to all schemes, except that only \sysname is able to utilize it.
Table~\ref{tab:computation_time} provides a breakdown
of the computation time for each scheme.

\item
\parai{Satisfied demand:} We focus on the percentage of
demand satisfied by a TE scheme in a practical \textit{online}
setting~\cite{ncflow}, accounting for the delay in TE control.
This means that the current flow allocation will persist
until the TE scheme finishes computing a new allocation.
We note that the satisfied demand only normalizes the total flow
with respect to the total demand, making it an appropriate metric
for evaluating schemes that optimize the total flow.

\item
\parai{Offline satisfied demand:} We also present the satisfied demand
calculated in an idealized \textit{offline} setting
in \S\ref{sec:offline-performance}, where
TE schemes are assumed to complete flow allocation instantaneously.
This metric eliminates the impact of delay on TE control
and focuses solely on the allocation quality.

\end{itemize}

\subsection{\sysname vs. the state of the art}
\label{sec:comparison-baselines}

Figure~\ref{fig:basic} compares \sysname against the state-of-the-art schemes
on four network topologies.
Although \sysname is not designed for small topologies such as SWAN
and UsCarrier, where an LP-solver can also quickly find the optimal 
allocation, we include the results to demonstrate the trend
(note that the computation time in Figure~\ref{fig:basic_t} is in log scale).
As the network size increases, we observe that \sysname demonstrates
scalable performance precisely as intended, requiring less than 1 second of 
computation time while allocating comparable or higher demand on Kdl and ASN.
On ASN, for instance, \sysname achieves 197--625$\times$ speedups
relative to the baselines (LP-all is not viable)
while satisfying 6--32\% more demand.

\parab{Small topologies (SWAN and UsCarrier).}
All the evaluated schemes can compute flow allocation on SWAN and UsCarrier 
within seconds, e.g., LP-all takes less than 1 second to determine
the optimal allocation, eliminating the need for TE acceleration schemes.
Nonetheless, we observe that when NCFlow is applied to UsCarrier,
it can only meet 72.6\% of the demand (vs. 96.2\% for LP-all).
In contrast to its suboptimal performance,
\sysname retains a demand of 92.6\% that is close enough to the optimal.

\parab{Kdl.} On the larger Kdl topology with 754 nodes and 1,790 edges,
\sysname takes only 0.95 seconds on average to complete each flow allocation,
which is 7$\times$ faster than NCFlow, 13$\times$ faster than POP, 
27$\times$ faster than LP-top, and 616$\times$ faster than LP-all.
Meanwhile, \sysname satisfies 90.6\% of the demand, nearly the same
as the best-performing scheme LP-top (with a difference of only 0.14\%).
Among the other schemes, LP-all requires over 585 seconds for computation,
exceeding the 5-minute time budget and forcing it to
reuse stale routes from several intervals ago.
As a result, LP-all only allocates 87.2\% of the demand despite its ability
to find the optimal solution if granted unlimited run time.
NCFlow and POP, on the other hand, produce flow allocations quickly
as intended, yet they only satisfy
63.8\% and 87.7\% of the demand, respectively.

\begin{figure}[t]
\centering
\begin{subfigure}[t]{\columnwidth}
\centering
\includegraphics[width=0.9\textwidth]{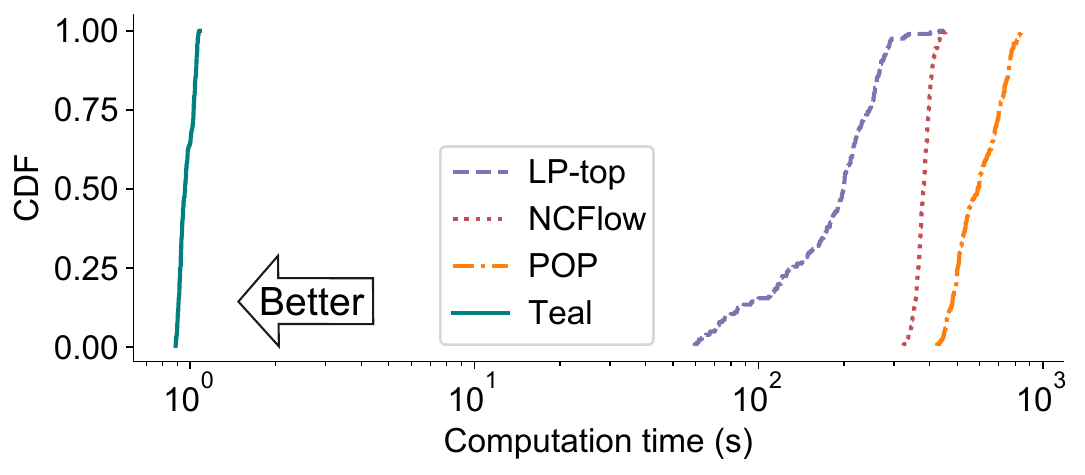}
\vspace{-2pt}
\caption{CDF of computation time on ASN}
\vspace{8pt}
\label{fig:basic_t_asn}
\end{subfigure}
\begin{subfigure}[t]{\columnwidth}
\centering
\includegraphics[width=0.9\textwidth]{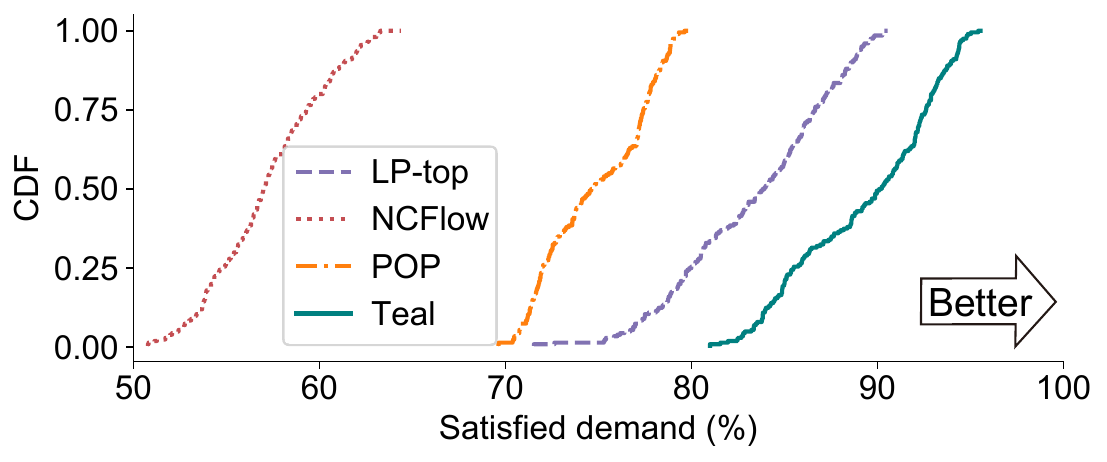}
\vspace{-2pt}
\caption{CDF of satisfied demand on ASN}
\label{fig:basic_d_asn}
\end{subfigure}
\caption{CDFs for the computation time and satisfied demand of schemes on ASN.
\sysname outperforms the baselines on both dimensions across
nearly all percentiles.}
\label{fig:cdf_asn}
\end{figure}

\parab{ASN.}
On the largest topology of ASN with 1,739 nodes and 8,558 edges,
\sysname achieves a more remarkable speedup relative to the other schemes.
With an average computation time of 0.97 seconds,
\sysname is 394$\times$ faster than NCFlow, 625$\times$ faster than POP,
and 197$\times$ faster than LP-top.
LP-all is impractical to run on ASN due to its incredibly
slow computation time of up to 5.5 hours per allocation
(4 orders of magnitude slower than \sysname), as well as the
memory errors incurred.
Not only does \sysname attain substantial acceleration,
it also allocates the most demand on average (89.4\%),
surpassing LP-top by 6\%, POP by 14.5\%, and NCFlow by 32\%.

Figure~\ref{fig:cdf_asn} zooms in on the performance of schemes
on ASN as CDF curves.
In Figure~\ref{fig:basic_t_asn}, we observe
that \sysname's computation time remains highly stable
across the tested traffic matrices,
staying within 0.89--1.08 seconds at all percentiles.
This remarkable stability can be attributed to the fact that
\sysname performs exactly one forward pass on its neural networks,
followed by precisely five iterations of ADMM (on this topology).
Thus,
the amount of computation (measured in floating-point operations) is
independent of the values in the input traffic matrix.
By contrast, the computation times of POP, NCFlow, and LP-top fluctuate
between 60--839 seconds, e.g., 257--730$\times$ slower than \sysname at the 90th
percentile.
The reason for this variability is that the LP solvers
employed in these methods have a stopping criterion that is
affected by problem-specific factors,
such as the ratio between traffic demands and link capacities.
Meanwhile, NCFlow involves nontrivial consolidation of the subproblem results,
and needs to iterate between LP optimization and consolidation
until a predefined accuracy threshold is reached.

Figure~\ref{fig:basic_d_asn} shows that \sysname achieves the highest flow
allocation across all percentiles.
Compared with NCFlow, POP, and LP-top, \sysname's satisfied demand
is 6--33\% higher at the median, and 5--33\% higher at the 90th percentile.
We believe that \sysname's robust performance makes it a compelling
choice for production TE systems.

\subsection{Reacting to link failures}
\label{sec:eval_fail}

\sysname efficiently solves TE optimization within a second, even for
large topologies with thousands of nodes such as ASN.
The real-time computation allows \sysname to
react promptly to link failures~\cite{ncflow}, as
\sysname can quickly recompute flow allocation on the altered
topology (with failed links having zero capacities).

\begin{figure}[t]
\centering
\includegraphics[height=90pt]{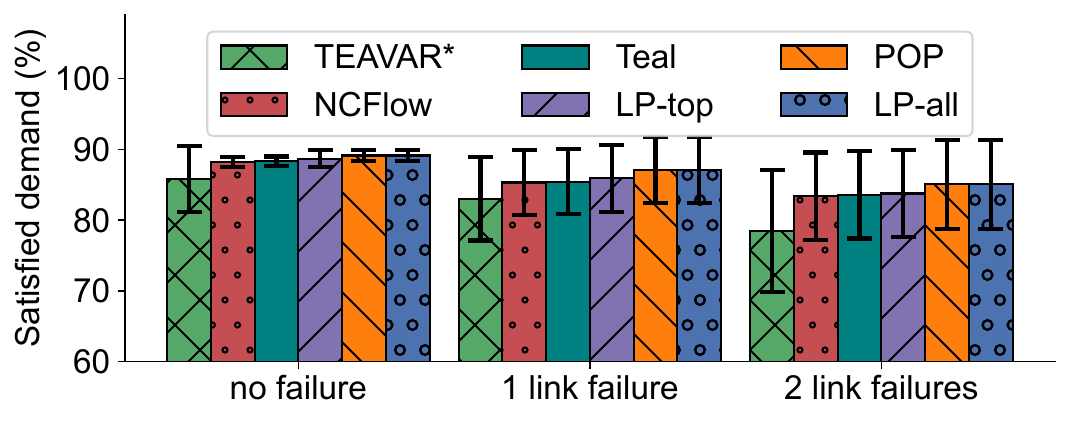}
\vspace{-4pt}
\caption{Satisfied demand of TE schemes in the presence of zero, one, or
two link failures on the small B4 network.
\sysname consistently outperforms TEAVAR$^*$ while
remaining on par with the other schemes.}
\label{fig:failure}
\vspace{-4pt}
\end{figure}

Although TEAVAR$^*$ is designed explicitly for fault tolerance under link failures, 
it is only viable on the small B4 network due to its significant computational
overhead. Therefore, we first evaluate all TE schemes on B4 and plot their
immediate allocation performance after the introduction of one or two link 
failures, with no link failures serving as a baseline.
Figure~\ref{fig:failure} shows that as the number of link failures increases,
the demand satisfied by all the tested schemes declines as expected.
Nevertheless, \sysname consistently outperforms
TEAVAR$^*$ by 2.4--5.1\%, while being statistically indistinguishable from
the other schemes.
It can be seen that in preparation for potential link failures, TEAVAR$^*$
has sacrificed link utilization for higher availability.
In contrast, we concur with NCFlow's viewpoint~\cite{ncflow}:
the performance decline during transient link failures can be
compensated through rapid recomputation.

\begin{figure}[t]
\centering
\includegraphics[height=90pt]{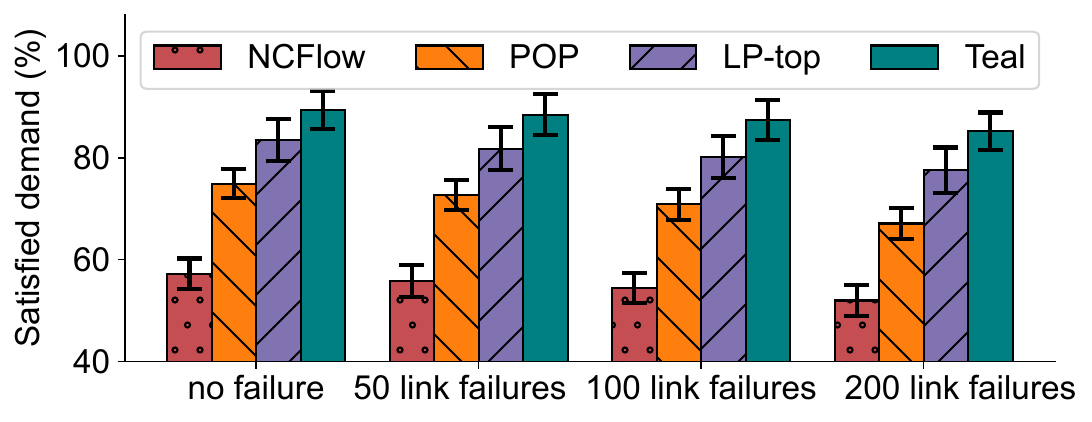}
\vspace{-4pt}
\caption{Satisfied demand of TE schemes in the presence of zero, 50, 100, or 200
link failures on the large ASN network. Through fast recomputation,
\sysname effectively minimizes the duration impacted by link failures
and thus preserves flow allocation performance.}
\label{fig:failure_asn}
\end{figure}

In practice, massive inter-datacenter link failures are very rare.
Even with fiber link failures, they do not usually translate to loss of
capacity on an inter-datacenter link (unless due to a fiber cut)~\cite{radwan}.
As a stress test, however,
we evaluate the extreme failure scenarios depicted in prior
work~\cite{arrow} and artificially inject 50, 100, and 200 link failures
to the ASN network. Figure~\ref{fig:failure_asn} presents the results
for the only 4 schemes feasible on ASN.
From the figure, we observe a similar trend across a variety of link failures: 
\sysname is able to route substantially more traffic demand than the baseline TE 
schemes, and the ranking is consistent with their run times shown in
\S\ref{sec:comparison-baselines}.
Specifically, \sysname (with a run time less than 1 second) satisfies 6--8\% more
demand than LP-top (191 s),
15--18\% more demand than POP (382 s), and
32--33\% more demand than NCFlow (606 s).
The reason is that the baseline TE schemes take significantly
longer than \sysname to recompute new allocations upon link failures.
As a result,
thousands of traffic flows have traversed the failed links and been dropped
during the computation of rerouting strategies.

It is noteworthy that \sysname achieves the above performance without
having to retrain its neural network models, showcasing its generality
across (transient) link failures.
In the less common event of persistent link failures
or planned network upgrades such as new network nodes or links,
we anticipate having sufficient time to retrain \sysname within 6--10 hours
and recover its allocation performance on the updated topology.

\subsection{Robustness to demand changes}
\label{sec:robustness}

\begin{figure}[t]
\centering
\begin{subfigure}[b]{0.49\columnwidth}
\centering
\includegraphics[height=110pt]{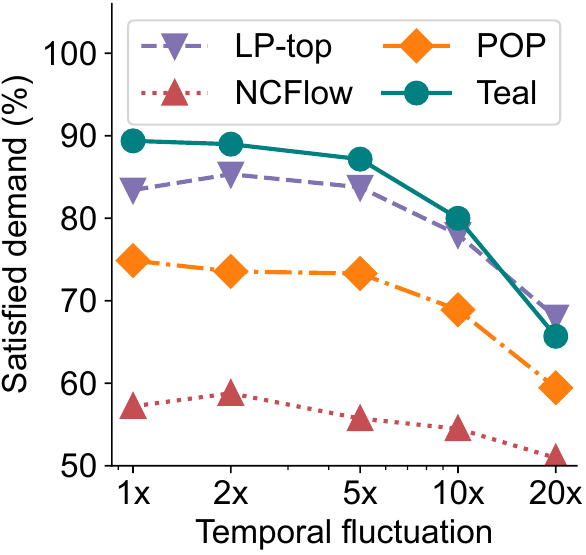}
\vspace{-10pt}
\caption{Temporal fluctuations}
\label{fig:robust_temporal}    
\end{subfigure}
\hfill
\begin{subfigure}[b]{0.49\columnwidth}
\centering
\includegraphics[height=110pt]{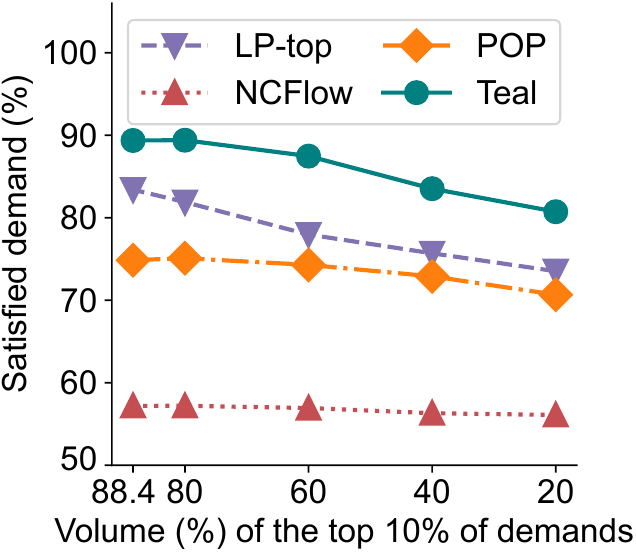}
\vspace{-10pt}
\caption{Spatial distributions}
\label{fig:robust_spatial}
\end{subfigure}
\vspace{-3pt}
\caption{Satisfied demand with temporal and spatial changes.}%
\label{fig:robust}
\end{figure}

We assess the robustness and generalization of \sysname concerning
temporal and spatial variations in traffic demands.
We do not specifically address massively unforeseen demands
because these scenarios pose a lesser concern in cloud WANs,
in contrast to ISP WANs~\cite{wang2006cope},
due to the mitigating effects of bandwidth brokering and TE feedback loop.
Nevertheless, in the event that \sysname's performance deteriorates
(e.g., in the face of exceptional demand changes), we may concurrently execute an
additional TE scheme, such as LP-top, to compute traffic allocation.
We can then seamlessly fall back to it if
it consistently yields superior solutions than \sysname.

\parab{Temporal fluctuation.} We introduce various temporal fluctuations
to the traffic matrices.
For each demand, we calculate the variance in its changes between
consecutive time slots, and multiply it by a factor of 2, 5, 10, and 20
to instantiate the variance of a zero-mean normal distribution.
Next, we randomly draw a sample from this normal distribution and add it
to each demand in every time slot.
Figure~\ref{fig:robust_temporal} shows that almost all the evaluated schemes
handle small fluctuations (2$\times$ and 5$\times$) effectively, but
their performance declines noticeably as the fluctuations escalate to
10$\times$ and 20$\times$.
Under 10$\times$ fluctuation, \sysname remains the top performer among all
schemes. Under the most severe 20$\times$ fluctuation, \sysname starts
to lag behind LP-top (by 2.3\%) due to not seeing the pattern
during training, but still outperforms
NCFlow and POP by 6--15\%.

\parab{Spatial distribution.}
We redistribute traffic demands across node pairs to simulate changes in
their spatial distribution. Specifically, we reassign the top 10\% of demands,
which originally account for 88.4\% of the total volume, such that
they constitute 80\%, 60\%, 40\%, and 20\% instead.
As shown in Figure~\ref{fig:robust_spatial}, \sysname consistently
satisfies the most demand across all spatial distributions.
LP-top's performance is reduced by $\sim$10\%, as its heuristic
is inherently reliant on the heavy-tailed demand distribution.

\subsection{\sysname under different objectives}
\label{sec:eval-objectives}
In this section, we evaluate the applicability of \sysname by retraining it
for two different TE objectives:
\textit{(i)} minimizing the max link utilization
(MLU)~\cite{fleischer2000approximating,learning-to-route},
and \textit{(ii)} maximizing the latency-penalized total flow~\cite{mate}.
Recall from \S\ref{sec:marl} that this is possible due to
the flexibility of \sysname's RL component with respect to the objective
to optimize.
Although the above objectives can be transformed into a form
compatible with ADMM similarly (as shown in Appendix~\ref{sec:admm-details}),
we opt to omit ADMM in these experiments as the neural network model
already exhibits satisfactory performance.
We compare \sysname against LP-all and LP-top since they are directly 
applicable to the new objectives, which are also linear.
However, adapting the codebases of NCFlow and POP to other objectives
is challenging, so they are not included in this section.

\begin{figure}[t]
\centering
\begin{subfigure}[b]{0.49\columnwidth}
\centering
\includegraphics[height=110pt]{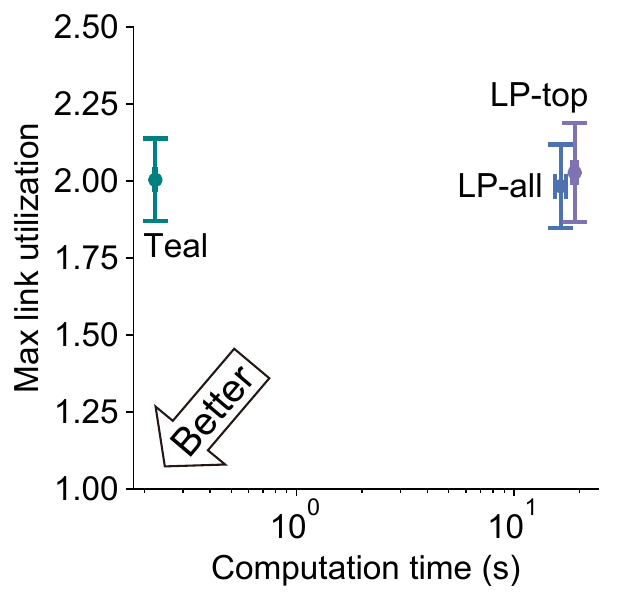}
\vspace{-10pt}
\caption{Kdl}
\label{fig:mlu_kdl}
\end{subfigure}
\hfill
\begin{subfigure}[b]{0.49\columnwidth}
\centering
\includegraphics[height=110pt]{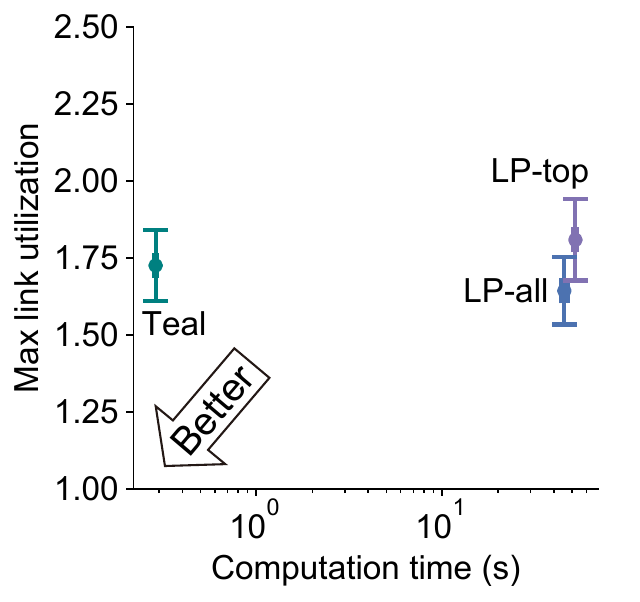}
\vspace{-10pt}
\caption{ASN}
\label{fig:mlu_asn}
\end{subfigure}
\vspace{-3pt}
\caption{Performance of \sysname and baselines under the TE objective
of minimizing max link utilization (MLU). All schemes attain
comparable MLU, but \sysname is 17--36$\times$ faster.}
\label{fig:mlu}
\end{figure}

\begin{figure}[t]
\centering
\begin{subfigure}[b]{0.49\columnwidth}
\centering
\includegraphics[height=110pt]{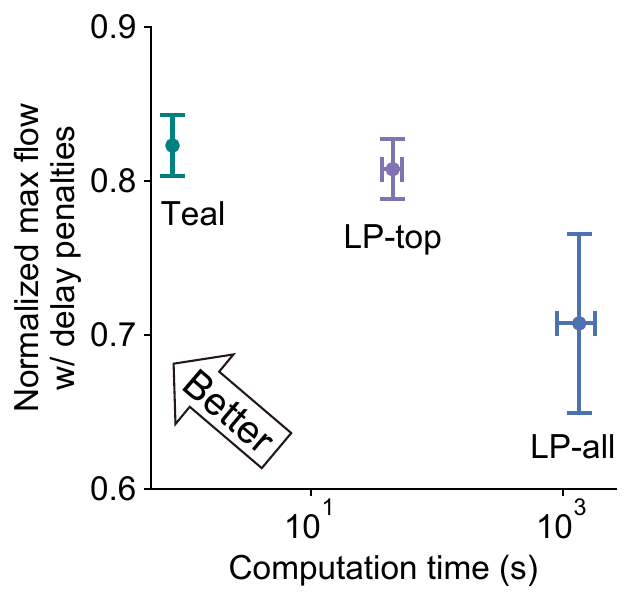}
\vspace{-10pt}
\caption{Kdl}
\label{fig:latency_kdl}
\end{subfigure}
\hfill
\begin{subfigure}[b]{0.49\columnwidth}
\centering
\includegraphics[height=110pt]{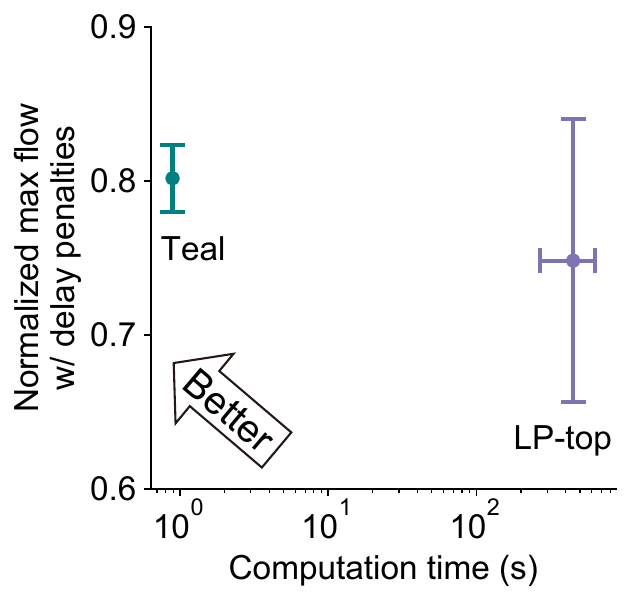}
\vspace{-10pt}
\caption{ASN}
\label{fig:latency_asn}
\end{subfigure}
\vspace{-3pt}
\caption{Performance of \sysname and baselines under the TE objective
of maximizing the total flow with delay penalties (LP-all is not
feasible on ASN). \sysname achieves the best allocation performance
while being 26--718$\times$ faster.}
\label{fig:latency}
\end{figure}

\parab{Max link utilization (MLU).} Figure~\ref{fig:mlu} shows
that all three schemes yield comparable MLUs, with no 
statistically significant differences.
However, \sysname finds a solution within only 0.22--0.29 seconds
when minimizing MLU,
whereas LP-all and LP-top require 73--85$\times$ longer on Kdl
and 158--181$\times$ longer on ASN.
Additionally, we observe two interesting phenomena.
First, LP-all and LP-top optimize MLU faster than the total flow,
presumably because minimizing MLU is ``easier'' and requires
fewer iterations for convergence in Gurobi.
Second, LP-all runs slightly faster than LP-top. The reason
is that the top 10\% of demands vary over time and thus require LP-top to constantly 
rebuild its model in Gurobi (see Table~\ref{tab:computation_time})
incurring additional computational overhead.

\parab{Latency-penalized total flow.} 
As shown in Figure~\ref{fig:latency}, \sysname's solution quality
is comparable to or higher than the best-performing scheme LP-top,
while being 56--505$\times$ faster in speed.
LP-all is not feasible on ASN when optimizing for this objective,
while being several orders of magnitude (1693$\times$) slower than \sysname on Kdl.

\subsection{Offline TE performance}
\label{sec:offline-performance}

\begin{figure}[t]
\centering
\begin{subfigure}[t]{0.49\columnwidth}
\centering
\includegraphics[height=110pt]{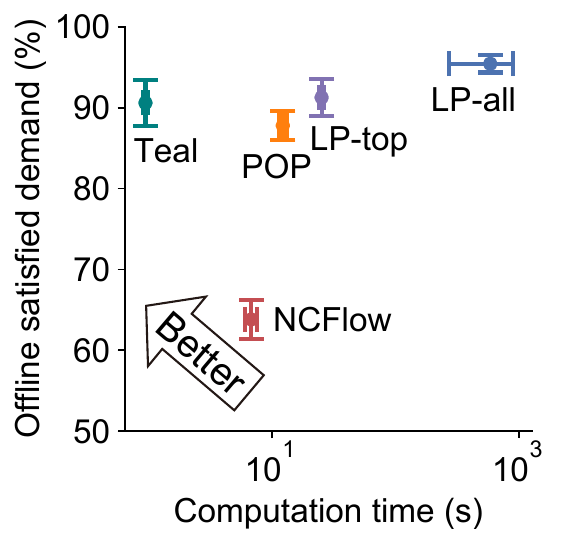}
\caption{Kdl}
\label{fig:offline_kdl}
\end{subfigure}
\hfill
\begin{subfigure}[t]{0.49\columnwidth}
\centering
\includegraphics[height=110pt]{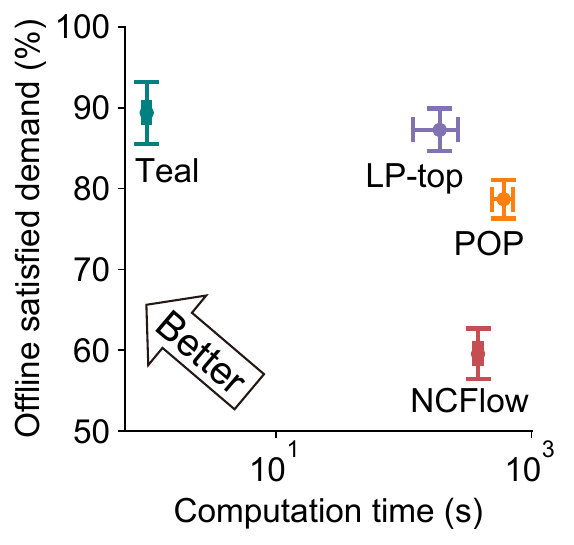}
\caption{ASN}
\label{fig:offline_asn}
\end{subfigure}
\vspace{-3pt}
\caption{Comparing the offline satisfied demand (defined in
\S\ref{sec:methodology}) of \sysname with baselines
(LP-all is not feasible on ASN).
\sysname's allocation quality remains close to optimal even when the computational delay is not taken into account.}
\label{fig:offline}
\vspace{-3pt}
\end{figure}

To determine the extent to which \sysname's performance benefits arise from
its fast and scalable computation, we evaluate all schemes under the offline
setting described in \S\ref{sec:methodology}.

On Kdl, while LP-all requires over 500 seconds to compute each 
flow allocation and exceeds the allotted time budget,
its output allocation is optimal and serves as a benchmark.
\sysname falls short of the optimal allocation by 4.8\% with respect to
the offline
satisfied demand, but remains within 0.7\% of the best feasible scheme
LP-top, and outperforms NCFlow by a significant margin of 27\% and POP
by 2.8\%, respectively.
On ASN, \sysname and LP-top achieve a similar level of offline satisfied demand,
which is 30\% higher than NCFlow and 11\% higher than POP. 

These findings suggest that even when the computational delay in TE control
is not taken into account,
\sysname's flow allocation quality is still close to optimal.
However, we note the caveat that \sysname achieves this by essentially
``overfitting'' the WAN topology, link capacities, and demand distribution.
For example, when faced with significant out-of-distribution demands (as shown in
\S\ref{sec:robustness}), the knowledge learned by \sysname may struggle
to apply and maintain the allocation performance.

\subsection{Ablation study of \sysname}
\label{sec:eval_interpret}

We perform an ablation study to assess the impact of \sysname's
key features on its overall performance.

\begin{figure}[t]
\centering
\begin{subfigure}[b]{0.50\columnwidth}
\centering
\includegraphics[height=100pt]
{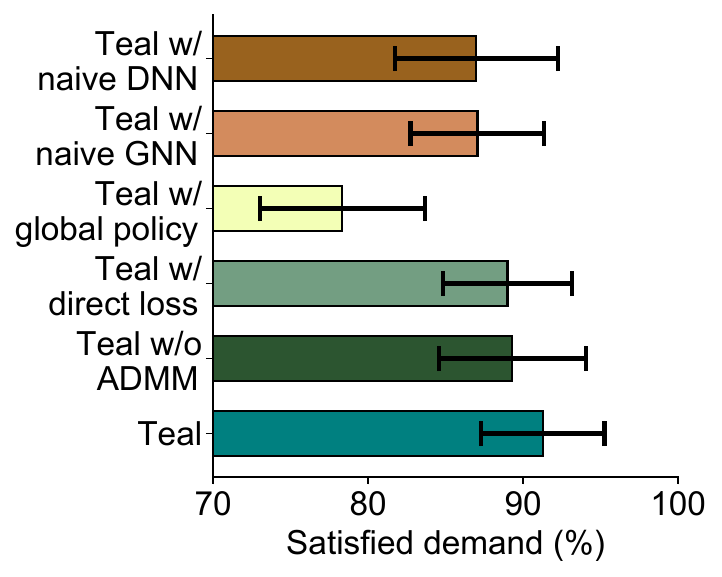}
\caption{SWAN}
\label{fig:ablation_swan}
\end{subfigure}
\begin{subfigure}[b]{0.48\columnwidth}
\centering
\includegraphics[height=100pt]
{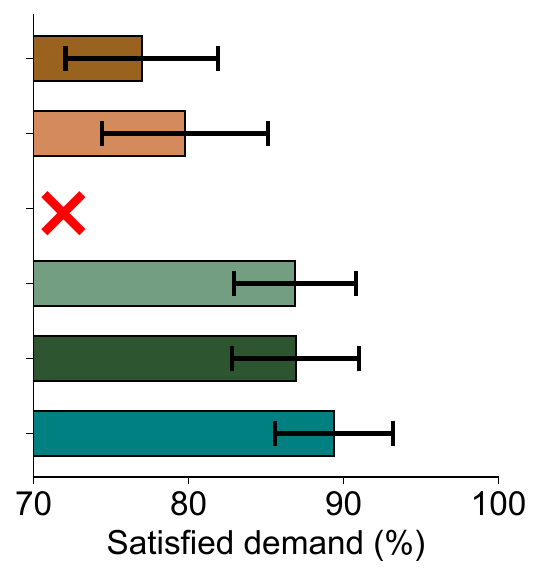}
\caption{ASN}
\label{fig:ablation_asn}
\end{subfigure}
\caption{Ablation study of \sysname's key features in its
FlowGNN, multi-agent RL, and ADMM components. Each feature proves useful for
\sysname's allocation performance.}
\label{fig:ablation}
\end{figure}

\parab{Design of FlowGNN.}
We devise two alternative designs for FlowGNN.
The first design, called ``\sysname w/ naive DNN,''
employs a 6-layer fully-connected neural network that
directly takes a traffic matrix as input and outputs traffic splits.
The second, called ``\sysname w/ naive GNN,''
models the WAN topology as a GNN directly,
with each node in the GNN representing a network site in the WAN
for feature learning.
This design enables information exchange among neighboring nodes in the WAN,
but fails to capture the relationship between edges and paths,
or network flows at the path level.
Figure~\ref{fig:ablation} reveals that compared with \sysname,
these two variants allocate 4.2--4.3\% less demand on SWAN
and 9.6--12.4\% on ASN, accentuating the importance of FlowGNN.

\parab{Processing demands independently.}
In contrast to \sysname's independent allocation of each demand,
an alternate approach described in \S\ref{sec:marl} involves processing
all demands at once using a ``gigantic policy network.''
This variant, referred to as ``\sysname w/ global policy,''
is not feasible for large networks such as ASN due to memory errors.
On the smaller SWAN network, it allocates 12.9\% less demand on average
compared with the full-fledged \sysname (Figure~\ref{fig:ablation_swan}).

\parab{Use of multi-agent RL.}
As discussed in \S\ref{sec:marl}, \sysname's multi-agent RL policy
can be replaced with direct loss minimization if
a non-differentiable TE objective is approximated by a surrogate loss.
For the total (feasible) flow, we define a surrogate loss as the total
intended flow (ignoring link capacities) minus the total overused
capacities (formal definition is in Appendix~\ref{sec:te-formulation}).
This variant, denoted as ``\sysname w/ direct loss,'' allocates
2.3--2.5\% less demand on average (Figure~\ref{fig:ablation}),
presumably due to the approximation error in the surrogate loss.
Moreover, \sysname's multi-agent RL policy may optimize a flexible array
of TE objectives (\S\ref{sec:eval-objectives}), while it is nontrivial to
identify a good surrogate loss for a new objective.

\parab{Fine-tuning with ADMM.}
Removing ADMM from \sysname's pipeline results in
a decline of 2--2.5\% in the satisfied demand, as indicated by
``\sysname w/o ADMM'' (Figure~\ref{fig:ablation}).
Although the impact is tolerable,
ADMM is a transparent optimization algorithm that strictly
reduces constraint violations when applied,
fine-tuning the solution with negligible run-time overhead.
We believe these properties make ADMM a desirable option for WAN operators.

\begin{figure}[t]
\centering
\begin{subfigure}[t]{0.32\columnwidth}
\centering
\includegraphics[width=0.95\textwidth]{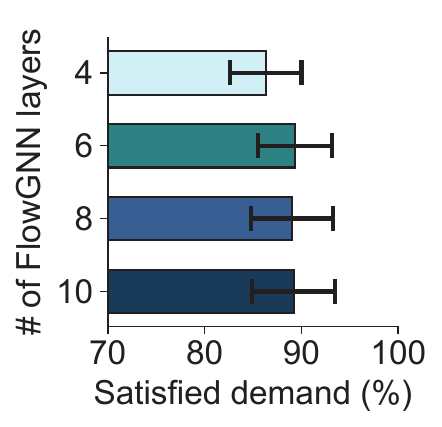}
\caption{FlowGNN layers}
\label{fig:sensitivity_flowGNN}
\end{subfigure}
\hfill
\begin{subfigure}[t]{0.32\columnwidth}
\centering
\includegraphics[width=0.95\textwidth]{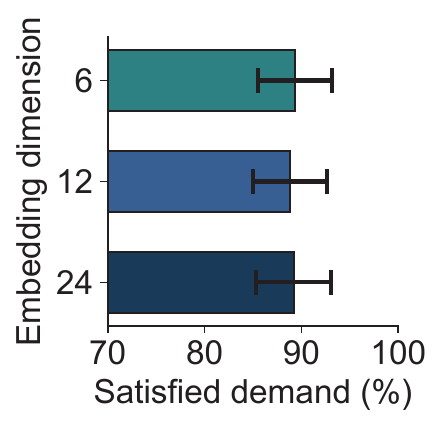}
\caption{Embedding sizes}
\label{fig:sensitivity_neuro}
\end{subfigure}
\hfill
\begin{subfigure}[t]{0.32\columnwidth}
\centering
\includegraphics[width=0.95\textwidth]{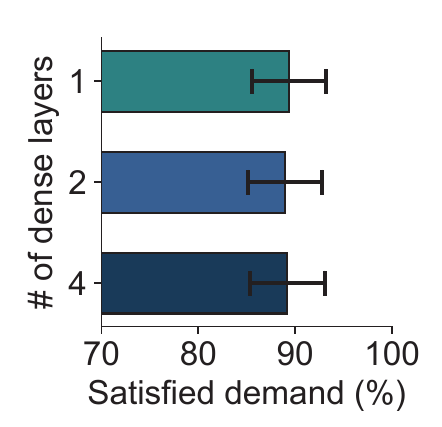}
\caption{Dense layers}
\label{fig:sensitivity_dense}
\end{subfigure}
\vspace{-3pt}
\caption{Sensitivity analysis of \sysname's hyperparameters.}%
\label{fig:sensitivity}
\end{figure}

\parab{Sensitivity analysis.}
We further conduct an analysis on the sensitivity of \sysname's performance
with respect to its hyperparameters. While the analysis is performed
on the ASN topology, similar results are observed across other topologies.
Figure~\ref{fig:sensitivity_flowGNN} depicts the impact of varying the
number of layers in FlowGNN.
As the number of layers increases from 4 to 6, \sysname's satisfied demand rises
from 86.3\% to 89.4\%, with diminishing returns beyond 6 layers.
Additionally, we explore different embedding dimensions in FlowGNN.
Instead of having a final output embedding of 6 elements
(by incrementing the embedding dimension by one in each FlowGNN layer),
we also test
higher final embedding dimensions such as 12 and 24.
However, the improvements achieved with higher embedding dimensions are marginal,
as indicated by Figure~\ref{fig:sensitivity_neuro}.

In Figure~\ref{fig:sensitivity_dense}, we vary the 
number of fully-connected (dense) layers in the policy network of \sysname's multi-agent RL,
and observe little difference in the allocation quality. 
This outcome aligns with our expectations since FlowGNN already 
captures the complex capacity-demand relationship within its architecture.
As a result, multi-agent RL primarily focuses on the task of transforming 
embeddings into split ratios, requiring only a lightweight structure.

\subsection{Visualization of flow embeddings}
\label{sec:visualization}

\begin{figure}[t]
\centering
\includegraphics[width=0.28\textwidth]{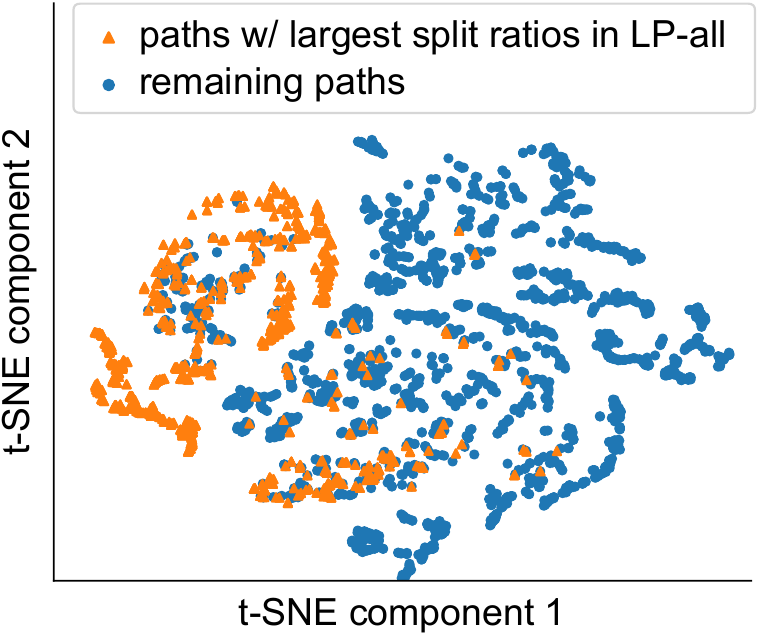}
\caption{Visualization of embeddings in FlowGNN.}%
\label{fig:visualization}
\end{figure}

To gain insights into the behaviors of \sysname,
we visualize the flow embeddings learned by FlowGNN for
the SWAN topology through a technique known as t-SNE
(t-distributed stochastic neighbor embedding)~\cite{tsne}.
The resulting visualization is shown in Figure~\ref{fig:visualization},
where the flow embeddings are projected onto a 2-dimensional space by t-SNE.
We color-code each flow embedding based on whether
its corresponding path is supposed to be ``busy'' in an optimal scenario,
i.e., it is assigned the largest split ratio among
the preconfigured paths in an optimal allocation generated by LP-all.

From the visualization, we observe an orange cluster of busy paths,
which is a useful indicator for the subsequent allocation task
as the downstream policy network can be trained to separate
this cluster from the remaining paths and allocate more traffic to the
paths desired to be busy, thereby mimicking the optimal solution provided by LP-all.
In other words, this cluster indicates that FlowGNN
has roughly captured path congestion within the network in its learned embeddings.

However, it is noteworthy that Figure~\ref{fig:visualization}
also contains a small number of outliers.
This is because TE optimization can yield multiple
optimal (or near-optimal) solutions. As a result,
the solution generated by \sysname might not be identical to that produced by LP-all, leading to the discrepancy between the two approaches.

%% file: relwork.tex
\section{Related Work}
\label{sec:relwork}
TE has been an integral part of 
service and cloud provider networks.
Network operators have leveraged TE to maximize network utilization,
guarantee fairness among flows,
and prevent link overutilization.
While ISPs used switch-native protocols (\eg MPLS, OSPF) to engineer traffic
in their networks~\cite{te-mpls, te-ospf},
cloud providers implemented centralized software-defined
TE systems to explicitly optimize 
for desirable network characteristics, such as low latency,
high throughput, and failure resilience~\cite{swan, b4, b4after,
ffc, blastshield, espresso, teavar}. In this section,
we place \sysname in the context of related work on cloud WAN TE.

\myparab{Intra-WAN traffic engineering.}
In the last decade, large cloud providers have deployed
SDN-based centralized TE in their planet-scale WANs to 
allocate traffic between datacenters~\cite{b4, swan, blastshield}.
Centralized TE systems formulate the
problem of allocating traffic in the WAN as an optimization
problem and periodically solve it to 
compute flow allocations. Due to the increase in
the scale of WANs and traffic matrices, the time required to solve the optimization problem has become a bottleneck in the 
TE control loop. Researchers have proposed
techniques that solve the TE optimization on smaller subsets
of the problem and combine the solutions to 
compute traffic allocations for the global graph~\cite{ncflow, pop}.
\sysname tackles the scalability challenges faced by modern
intra-WAN TE controllers using a learning-based approach. 

\myparab{Inter-WAN traffic engineering.}
Cloud providers engineer traffic at the edge of their networks by allocating demands on the links between the cloud and ISPs. Recent work has shown the role of engineering inter-WAN traffic for performance improvement and cost reduction~\cite{cascara, edgefabric}.
In contrast, \sysname focuses on intra-WAN TE.

\myparab{ML for traffic engineering.}
Deep learning and broader ML techniques have seen applications
in a range of classical networking problems, including
adaptive video streaming~\cite{pensieve,puffer},
TCP congestion control~\cite{aurora,pantheon}, and
traffic demand prediction~\cite{kumar2019cloud,mallick2020dynamic}.
Recently, researchers have begun leveraging ML to allocate traffic
in WANs~\cite{learning-to-route,learning-to-route2,dote},
focusing on learning to route under traffic uncertainty
and exploiting the predictive power of ML to improve allocation performance.
However, production WAN TE still heavily relies on separate components 
such as bandwidth brokers to provide the traffic matrix
for the next TE time step as input to the TE controller.
Other ML-based approaches to TE~\cite{te-rl, te-rl2, te-rl3, te-rl4, te-rl5}
operate under a variety of assumptions and do not apply to the 
acceleration of large-scale intra-WAN TE.
Our work demonstrates that learning-based approaches can significantly 
accelerate TE optimization while achieving near-optimal
allocation performance, addressing the increasing scale of TE optimization.

%% file: conclusion.tex
\section{Conclusion}
\label{sec:conclusion}
In this work, we demonstrate that deep learning is an effective tool for scaling
cloud WAN TE systems to large WAN topologies.
We develop \sysname, a learning-based TE scheme
that combines carefully designed, highly parallelizable
components---FlowGNN, multi-agent RL,
and ADMM---to allocate traffic in WANs.
\sysname computes near-optimal traffic allocations with substantial
acceleration over state-of-the-art TE schemes for large WAN topologies.

\parab{Ethics:} This work does not raise any ethical issues.

%% file: appendix.tex
\newpage
\appendix

\noindent{\LARGE \textbf{Appendices}}

\noindent{Appendices are supporting material that has not been
peer-reviewed.}

\section{TE optimization formulation}
\label{sec:te-formulation}

The goal of cloud WAN traffic engineering (TE) algorithms is to efficiently utilize the expensive network
resources between datacenters to achieve operator-defined performance goals, 
such as minimum latency, maximum throughput, and fairness between
customer traffic flows.

\parab{Network.}
We represent the WAN topology as a graph $G = (V, E, c)$,
where nodes ($V$) represent network sites (e.g., datacenters), 
edges ($E$) between the sites represent network links resulting from
long-haul fiber connectivity, and
$c: E\rightarrow \mathbb{R}^+$ assigns capacities to links.
Let $n = |V|$ denote the number of network sites. Each network
site can consist of either one or multiple aggregated routers.

\parab{Traffic demands.} The demand $d \in D$ between a pair of
network sites $s$ and $t$ in $G$ is the volume of network traffic
originating from $s$ that must be routed to $t$ within
a given duration of time. A separate component in the system
(such as a bandwidth broker) periodically gauges demands for the next
time interval (e.g., five minutes) based on the needs of various services,
historical demands, and bandwidth enforcement~\cite{b4}.
The gauged demand is considered fixed for the next time interval and provides
as input to the TE optimization. The TE algorithm computes allocations along network paths to meet the given demand~\cite{learning-to-route, ncflow}.

\parab{Network paths.}
The network traffic corresponding to a demand $d$ flows on
a set of preconfigured network paths $P_{d}$. These paths are 
precomputed by network operators (e.g., using the shortest paths) and serve as
input to the TE optimization. This version of TE optimization that allocates demands onto preconfigured paths as opposed to individual edges,
is known as the path formulation of TE, which is widely adopted in
production WANs~\cite{swan,b4,b4after,blastshield}.
Path formulation reduces the computational complexity of the TE optimization and also reduces the number of switch forwarding entries required to implement the traffic allocation.

\parab{Traffic allocations.} A traffic allocation $\mathcal{F}$ allocates a
demand $d \in D$ as flows across the assigned network paths $P_{d}$.
Therefore, $\mathcal{F}_d$ is a mapping from the path set $P_{d}$ to
non-negative split ratios,
i.e., $\mathcal{F}_{d}: P_{d} \rightarrow [0, 1]$, such
that $\mathcal{F}_{d}(p)$ is the fraction of traffic demand $d$ allocated
on path $p$. The traffic allocation in time interval $i$ is denoted as
$\mathcal{F}^{(i)}$. 

\parab{Constraints.}
For any demand $d \in D$, 
$\sum_{p \in P_d} \mathcal{F}_{d}(p) \leq 1$ 
is maintained such that we only allocate as much traffic as the demands.
Additionally, we constrain the allocations by $e \in E$, $c(e) \geq \sum_{p \ni e} \sum_{d \in D} \mathcal{F}_d(p) \cdot d$
to ensure that the traffic allocations do not exceed the capacity of network links.

\parab{TE objectives.}
The goal of TE algorithms can range from maximizing 
network throughput to minimizing latency, and previous work
has explored algorithms with a variety of TE objectives.
We show that \sysname can achieve near-optimal allocation with substantial
acceleration for different well-known TE objectives (\S\ref{sec:eval}).
In this section, we illustrate
the TE optimization problem using the maximum network flow objective since it
has been adopted by production TE systems~\cite{swan, blastshield}. The TE
optimization computes a routing policy $\mathcal{F}$ that satisfies the demand
and capacity constraints while maximizing the TE objective.
Equation~(\ref{eqn:network-opt})
summarizes our TE formulation:

\begin{equation}
\label{eqn:network-opt}
\begin{aligned}
\textrm{maximize } \quad & \sum_{d \in D} \sum_{p\in P_{d}} \mathcal{F}_{d}(p) \cdot d \\
\textrm{subject to } \quad
&\sum_{p \in P_{d}} \mathcal{F}_d(p) \leq 1, \forall d \in D\\
&\sum_{p \ni e} \sum_{d \in D} \mathcal{F}_d(p) \cdot d \leq c(e), \forall e \in E\\
& \mathcal{F}_{d}(p) \geq 0, \forall d \in D, \forall p \in P_{d}
\end{aligned}
\end{equation}

\parab{Surrogate loss.} 
The surrogate loss that approximates the
(non-differentiable) total feasible flow is defined as the
total flow intended to be routed (disregarding link capacities),
penalized by total link overutilization. Using the above notations,
the surrogate loss can be formally expressed as
\[\sum_{d \in D} \sum_{p\in P_{d}} \mathcal{F}_{d}(p) \cdot d - \sum_{e \in
E}\max(0, \sum_{p \ni e} \sum_{d \in D} \mathcal{F}_d(p) \cdot d - c(e)).\]

\vspace{10pt}
\section{COMA$^*$ details}
\label{sec:coma-details}

At a high level, COMA builds on the idea of counterfactual reasoning~\cite{wolpert2002optimal},
deducing the answer to a ``What if...'' question:
At the moment every agent is about to make a decision (\textit{action}),
what would be the difference in global reward if
\textit{only} one agent's action changes while the other agents'
actions remain fixed?
E.g., in the context of TE that aims to maximize the total flow,
our COMA$^*$ reasons about:
Compared with the current traffic allocations, how much would the total flow
differ if we only reallocate the flows of one demand while keeping the
allocations of the other demands unchanged? 
The performance difference measures the contribution of an agent's action to the overall
reward.
Specifically, the reward difference defines the ``advantage'' 
of the current joint action over the counterfactual baseline
(where only one agent tweaks its action). The advantage is heavily used
in this family of RL algorithms (known as actor-critics~\cite{konda1999actor})
to effectively reduce the variance in training.

At each time step when a new traffic matrix arrives
or any link capacity changes (e.g., due to a link failure), \sysname passes the
flow embeddings (stored in PathNodes of FlowGNN) for the same
demand to the RL agent $i$ designated to manage the demand. We define these
flow embeddings as the \textit{state} $s_i$ observed locally by agent $i$. Presented
only with the local view captured by $s_i$, agent $i$ makes an action
$a_i$, a vector of split ratios that describes the allocation of the agent's
managed demand.  Let $\pi_{\theta}$ denote the
policy network parameterized by $\theta$ shared by agents. Learning the
weights $\theta$ with gradient descent is known as policy
gradient~\cite{sutton1999policy}, which typically requires a stochastic form
$\pi_{\theta}(a_i|s_i)$ that represents the probability of outputting $a_i$
given $s_i$. Since allocations are deterministic in TE, a common way that
converts $\pi_{\theta}$ to stochastic is to have it output the mean and variance
of a Gaussian distribution. During training, actions are
sampled from the Gaussian distribution $a_i \sim \pi_{\theta}(\cdot |s_i)$,
whereas the mean value of the Gaussian is directly used as the action
during deployment.

We use $\mathbf{s}$ to denote the central state formed by all local states
$s_i$, and $\mathbf{a}$ to denote the joint action formed by all local actions
$a_i$.  A reward $R(\mathbf{s}, \mathbf{a})$, such as the total flow, is
available after all agents have made their decisions. To compute
the advantage $A_i(\mathbf{s}, \mathbf{a})$ when \textit{only} agent $i$
alters its action, COMA proposes to estimate the expected return,
namely a discounted sum of future rewards, obtained by taking the joint action
$\mathbf{a}$ in central state $\mathbf{s}$.
By comparison, our COMA$^*$ computes the expected return by leveraging the
``one-step'' nature of TE: an action (flow allocation) in TE does not 
impact the future states (traffic demands). Consequently, the expected return
effectively equals the reward $R(\mathbf{s}, \mathbf{a})$ obtained at a
single step. Moreover, suppose that agent $i$ varies its action to $a_i'$
while the other agents keep their current actions,
the new joint action---denoted as $(\mathbf{a}_{-i}, a_i')$---can be directly
evaluated by simulating its effect, i.e., we compute the TE objective
obtained if the new joint action were to be used.
Putting everything together, COMA$^*$ computes the advantage for agent $i$
as follows:
\begin{equation}
\label{eqn:advantage}
  A_i(\mathbf{s}, \mathbf{a}) = R(\mathbf{s}, \mathbf{a}) -
  \sum_{a_i'}\pi_{\theta}(a_i'|s_i)R(\mathbf{s}, (\mathbf{a}_{-i}, a_i')),
\end{equation}
where we perform Monte Carlo sampling to estimate the counterfactual
baseline, e.g., by drawing a number of random samples for 
$a_i' \sim \pi_{\theta}(\cdot |s_i)$. The gradient of $\theta$ is then given by
\begin{equation}
\label{eqn:policy-gradient}
  g = \mathbb{E}_\pi \left[ \sum_i A_i(\mathbf{s}, \mathbf{a})
                            \nabla_\theta \log \pi_\theta (a_i|s_i) \right],
\end{equation}
which is used for training the policy network with standard
policy gradient.
In practice, \sysname trains FlowGNN and the policy network of COMA$^*$
end to end, so $\theta$ represents all the parameters to learn in the end-to-end
model, backpropagating gradients from the policy network to FlowGNN.

\vspace{10pt}
\section{ADMM details}
\label{sec:admm-details}
In this section, we derive the ADMM iterates for the TE problem
in Equation~(\ref{eqn:network-opt}), reproduced here:
\begin{align}
\textrm{maximize } \quad & \sum_{d \in D} \sum_{p\in P_{d}} \mathcal{F}_{d}(p) \cdot d \nonumber\\
\textrm{subject to } \quad
&\sum_{p \in P_{d}} \mathcal{F}_d(p) \leq 1, \forall d \in D \label{eqn:1}\\
&\sum_{p \ni e} \sum_{d \in D} \mathcal{F}_d(p) \cdot d \leq c(e), \forall e \in E \label{eqn:2}\\
& \mathcal{F}_{d}(p) \geq 0, \forall d \in D, p \in P_{d}. \nonumber
\end{align}
In order to apply ADMM which requires a specific form to optimize, we must decouple the constraints in the original problem. As Constraint~(\ref{eqn:2}) couples the edge traffic across paths and demands, we introduce dummy variables $z_{pe}$ for each path $p$ (from in any demand $d\in D$), and edge $e \in p$.
We note that each path $p\in P_d$ uniquely stems from a particular demand $d$.
Then, we replace Constraint~(\ref{eqn:2}) with the following constraints:
\begin{align}
\sum_{p \ni e} z_{pe} &\leq c(e), \forall e \in E \label{eqn:3}\\
\mathcal{F}_d(p) \cdot d - z_{pe} &= 0, \forall d\in D,p\in P_d, e\in p. 
\label{eqn:4}
\end{align}
Finally, we add slack variables $s = (s_{1d},s_{3e})$, for all demands $d \in D$ and edges $e \in E$ respectively, to turn inequality in Constraint~(\ref{eqn:1}) and Constraint~(\ref{eqn:3}) into equality:
\begin{align}
\textrm{maximize } \quad & \sum_{d \in D} \sum_{p\in P_{d}} \mathcal{F}_{d}(p) \cdot d \nonumber\\
\textrm{subject to } \quad
&\sum_{p \in P_{d}} \mathcal{F}_d(p) + s_{1d} - 1 = 0, \forall d \in D \tag{\ref{eqn:1}}\\
&\sum_{p \ni e} z_{pe} + s_{3e} - c(e) = 0, \forall e \in E \tag{\ref{eqn:3}}\\
&\mathcal{F}_d(p) \cdot d - z_{pe}=0, \forall d\in D, p\in P_d,  e\in p \tag{\ref{eqn:4}}\\
& \mathcal{F}_{d}(p) \geq 0, \forall d \in D, p \in P_{d}. \nonumber
\end{align}
By introducing Lagrange multipliers $\lambda = (\lambda_1,\lambda_3,\lambda_4) \in \mathbb{R}^{|D|} \times \mathbb{R}^{|E|} \times \mathbb{R}^{|P||E|}$
and a penalty coefficient $\rho$,
the augment Lagrangian for this transformed problem becomes 
\begin{equation*}
\label{eqn:augmented-lagrangian}
\begin{aligned}
&\mathcal{L}_{\rho}(\mathcal{F}, z, s, \lambda)\\
&= -\sum_{d \in D} \sum_{p\in P_{d}} \mathcal{F}_{d}(p) \cdot d + \lambda G(\mathcal{F}, z, s) + \frac\rho2\|G(\mathcal{F},z,s)\|_2^2,
\end{aligned}
\end{equation*}
where $G(\mathcal{F},z,s) = (G_1,G_3,G_4)^\top$ and
\begin{align*}
G_{1d} &= \sum_{p \in P_d} \mathcal{F}_d(p) + s_{1d} - 1 \\
G_{3e} &= \sum_{p \ni e} z_{pe} + s_{3e} - c(e) \\
G_{4dpe} &= \mathcal{F}_d(p) \cdot d - z_{pe}.
\end{align*}
The ADMM iterates at step $k+1$ are then given by
\begin{align*}
\mathcal{F}^{k+1} &:= \argmin_{\mathcal{F}}\mathcal{L}_\rho(\mathcal{F}, z^{k},s^{k},\lambda^{k})\\
z^{k+1} &:= \argmin_{z}\mathcal{L}_\rho(\mathcal{F}^{k+1}, z,s^{k},\lambda^{k})\\
s^{k+1} &:= \argmin_{s}\mathcal{L}_\rho(\mathcal{F}^{k+1}, z^{k+1},s,\lambda^{k})\\
\lambda^{k+1} &:= \lambda^{k} + \rho\cdot G(\mathcal{F}^{k+1},z^{k+1},s^{k+1})
\end{align*}
with the initial iterates warm-started by the policy network.

\vspace{10pt}
\section{Topology details}
\label{sec:topology-details}

Table~\ref{tab:topologies-details} provides additional details about the network
topologies utilized in our study (SWAN is excluded from the table due to
containing private information). In general, as the size of the network topology 
increases, the average shortest-path length and network diameter tend to
become longer, with the exception of the ASN topology.
This can be attributed to the presence of star-shaped clusters (ASes) within
ASN. These clusters are interconnected, resulting in a strong connectivity 
at the cluster level.

\begin{table}[b]
  \centering
  \begin{tabular}{lll}
  \toprule
            & \begin{tabular}{@{}l@{}}Average shortest-\\path length \end{tabular} & Network diameter \\
  \midrule
  B4        & 2.3    & 5      \\
  UsCarrier & 12.1   & 35     \\
  Kdl       & 22.7   & 58     \\
  ASN       & 3.2    & 8      \\
  \bottomrule
  \end{tabular}
  \vspace{8pt}
  \caption{Additional topology details.}
  \label{tab:topologies-details}
\end{table}

\begin{figure}[t]
\centering
\includegraphics[height=100pt]{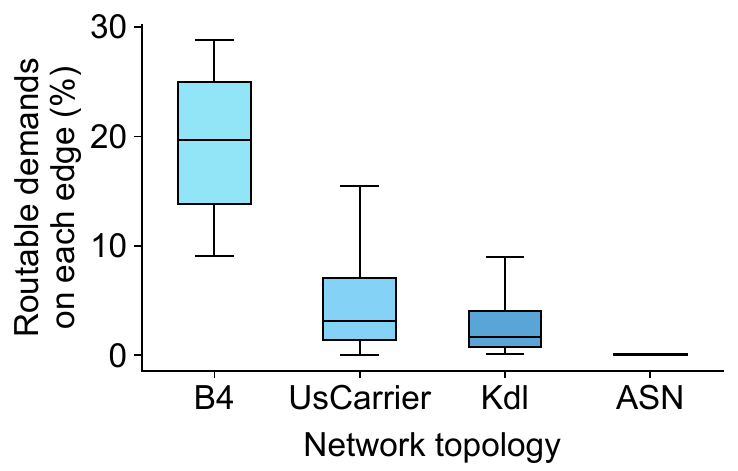}
\caption{Percentage of routable demands for each edge,
i.e., if (any of) the demand's preconfigured paths pass through the edge.}
\label{fig:topo}
\end{figure}

Meanwhile, we examine the percentage of demands (over the total number of
demands) that are routable on each edge, i.e., if the edge lies on at least
one of a demand's preconfigured paths, and plot the distributions
in Figure~\ref{fig:topo}.
This figure reveals that as the network grows in size,
each edge tends to serve a decreasing percentage of demands due to the
sparser distribution of demands.
Notably, the ASN topology exhibits an exceptionally low proportion
of routable demands on each edge due to its distinctive characteristics.

\vspace{10pt}
\section{TE performance over time}
\label{sec:allocation-over-time}

\begin{figure}[h]
\centering
\includegraphics[width=0.45\textwidth]{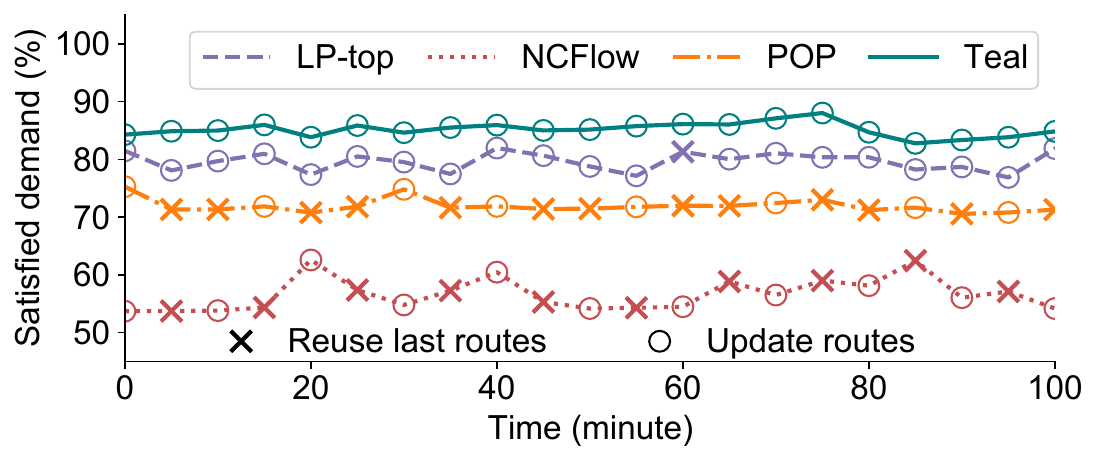}
\caption{Allocation performance of schemes in response to changing demands over time on ASN. 
\sysname consistently allocates the most demand in each time interval.}%
\label{fig:timeline_asn}
\end{figure}

We present the allocation performance of different schemes in response to
changing demands over time in Figure~\ref{fig:timeline_asn}.
With the run time fluctuating from a median of 200 s to the worst case 
of 450 s in Figure~\ref{fig:basic_t_asn}, LP-top only uses updated routes
at the end of the 5-minute interval and occasionally uses stale routes 
throughout the interval, leading to less demand satisfied.
We also observe that NCFlow and POP can only compute a new allocation for 
every other or every third traffic matrix, and using stale routes from 5 or 
10 minutes ago causes further performance degradation to the original
suboptimal traffic allocation.
In contrast, \sysname consistently allocates the most demand 
in each time interval.